\documentclass{iopart}
\def\eps{\varepsilon}
\def\fii{\varphi}
\def\d{ {\rm d} }

\def\Complex{ \mathbb{C} }
\def\scri{{\cal I}}
\def\s{\sigma^{(0)}}
\def\ds{\dot{\sigma}{}^{(0)}}
\def\sbar{\bar{\sigma}^{(0)}}
\def\dsbar{\dot{\bar{\sigma}}{}^{(0)}}
\def\b1{\bar{1}}
\newcommand{\SYMM}[1]{ \left( #1 \right)}
\newcommand{\ord}[1]{{\cal O}(#1)}
\newcommand{\qed}{\hfill $\Box$ \medskip}
\usepackage{iopams}
\usepackage{amssymb}
\usepackage{amsthm}
\usepackage{amsfonts}
\usepackage{fontenc}
\usepackage{graphicx}
\newtheorem{theo}{Theorem}[section]

\newtheorem{lemma}[theo]{Lemma}

\newcounter{mnotecount}[section]
\renewcommand{\themnotecount}{\thesection.\arabic{mnotecount}}

\newcommand{\mnote}[1]
{\protect{\stepcounter{mnotecount}}$^{\mbox{\footnotesize
$
\bullet$\themnotecount}}$ \marginpar{
\raggedright\tiny\em
$\!\!\!\!\!\!\,\bullet$\themnotecount: #1} }

\usepackage[notcite]{}

\newcommand{\NUMPARTS}[1]{\numparts \def\theequation{#1\arabic{eqnval}{\it \alph{equation}}}}
\newcommand{\ENDNUMPARTS}[1]{\endnumparts \def\theequation{#1\arabic{equation}}}

\begin{document}

\title[No periodic asymptotically flat solutions of the Einstein equations.]{On asymptotically flat solutions \\of Einstein's equations periodic in time\\ II. Spacetimes with scalar field sources}
\author{J Bi\v{c}\'ak$^{1,3}$, M Scholtz$^{1,3}$, P Tod$^{2,1}$}
~\\
\address{$^1$\,Institute of Theoretical Physics, Faculty of Mathematics and Physics, \\ Charles University, V Hole\v{s}ovi\v{c}k\'ach 2, 180 00 Prague 8, Czech Republic}~\\
\address{$^2$\, Mathematical Institute, Oxford OX1 3LB, UK}~\\
\address{$^3$\,Max Planck Institute for Gravitational Physics, Albert Einstein Institute, \\
Am M\"uhlenberg 1, 14476 Golm, Germany}
\eads{\mailto{bicak@mbox.troja.mff.cuni.cz}, \mailto{scholtzzz@gmail.com}, \mailto{paul.tod@sjc.ox.ac.uk}}

\begin{abstract}
We extend the work in our earlier article \cite{BST} to show that time-periodic, asymptotically-flat solutions of the Einstein equations analytic at $\scri$, whose source is one of a range of scalar-field models, are necessarily stationary. We also show that, for some of these scalar-field sources, in stationary, asymptotically-flat solutions analytic at $\scri$, the scalar field necessarily inherits the symmetry. To prove these results we investigate miscellaneous properties of massless and conformal scalar fields coupled to gravity, in particular Bondi mass and its loss.
\end{abstract}

\section{Introduction}
In this article, we continue the study begun in \cite{BST} (paper I) of asymptotically-flat solutions of Einstein's equations that
are periodic in time. In \cite{BST}, we showed that such
space-times, if either vacuum or electrovacuum and analytic
at $\scri$, are necessarily stationary near $\scri$. Here we
extend this result to space-times whose source is one of a
range of scalar field models.

In \cite{BST}, we also considered the problem of inheritance of
symmetry. This is the question of whether, if a space-time which is a solution of
Einstein's equations with some matter source has a symmetry, the matter source
necessarily has the same symmetry. For asymptotically-flat electrovacuum
space-times which are analytic near $\scri$ we showed that the symmetry is necessarily inherited.
For scalar field sources, we now obtain the same result in some cases but not in others.

The scalar fields we consider fall into two broad classes. The first class includes the
complex, massless Klein-Gordon (KG) field which satisfies the wave equation (\ref{KGEqPhys}) and has
energy-momentum tensor as in (\ref{KG-Tab}). Here we prove

\medskip
Theorem \ref{one}: \emph{A weakly-asymptotically-simple time-periodic solution of the Einstein-massless-KG field equations which is analytic in a neighbourhood of $\scri^-$
necessarily has a Killing vector which is time-like in the interior and extends to a translation on $\scri^-$.}

\medskip

\noindent It is also possible to include a potential for the scalar field, as in
subsection (\ref{ssPot}), and therefore to include a mass-term, and the above result will continue to hold subject to a weak condition on the potential.
Now one knows, for example from \cite{BW},
that there exist boson-star solutions of the Einstein-massive-KG system for which the metric is
static, spherically-symmetric and asymptotically-flat, while the complex scalar field takes the form $f(r)e^{i\omega t}$:
these solutions are genuinely periodic in time but not stationary, and the source does not inherit all
the symmetries of the metric. However, it is easy to see that these solutions are not analytic near $\scri$,
which is why they do not violate our result.

The other class of scalar fields contains what we shall call \emph{the conformal scalar field}, that is, it
satisfies the conformally-invariant wave equation (\ref{ConfInvWaveEq}). For simplicity,
we shall take the field to be real, though the formalism allows a complex field. In the real case, there is a conserved
energy-momentum tensor for such a source due originally to \cite{NP} (see also \cite{CCJ} and \cite{PR}),
given in (\ref{ConfCoupledTab}). This leads to a form of the Einstein equations (\ref{EinstEQS2})
studied in \cite{hub}, where it was shown that there
is a well-posed initial value problem. With suitable data on hyperboloidal surfaces extending to $\scri$, there exist asymptotically-flat solutions even with a regular point at $i^+$ \cite{hub}. Explicit static spherically-symmetric solutions were earlier given in \cite{bek1}, \cite{bek2}. Starting from the assumption of an asymptotically-flat solution of these Einstein equations, we may proceed as before,
with the corresponding result:

\medskip

Theorem \ref{two}: \emph{A weakly-asymptotically-simple time-periodic solution of the Einstein-conformal-scalar field equations which is analytic in a neighbourhood of $\scri^-$
necessarily has a Killing vector which is time-like in the interior and extends to a translation on $\scri^-$.}

\medskip

\noindent Turning to the question of inheritance, for the first class of fields we show (Theorem \ref{three}) that the only way
a stationary symmetry can fail to be inherited in the class of space-times under consideration
is if the (necessarily complex) scalar field has the
form $f(x^i)e^{i\omega t}$ in terms of comoving space-coordinates $x^i$ and time $t$. This is periodic and the previous result can be applied to deduce
that the symmetry is in fact inherited. For a complex conformal scalar field, the same argument can be used to show that there are no non--inheriting fields of this form but this is only a partial result as we cannot characterise the non-inheriting fields in the same way.

The plan of the article is as follows: in section 2 we review the Einstein-massless-KG equations and show how to formulate the \emph{conformal Einstein-massless-KG equations}, by which we mean the equations formulated for an unphysical, rescaled metric which correspond to the physical Einstein-massless-KG equations. This enables the equations to be extended to $\scri$. In section 3, we do the same thing for the Einstein-conformal-scalar equations, using the conserved enery-momentum tensor proposed in \cite{NP} (see also \cite{CCJ}). This energy-momentum tensor does not satisfy the Dominant Energy Condition, but we give some arguments why it might none-the-less lead to positive total energy. In section 4, we give expressions for the Bondi mass and Bondi mass-loss for both classes of scalar field sources. The Bondi mass-loss for the conformal scalar field is not manifestly positive (at $\scri^+$) but in the periodic case the average over a period is. In section 5, we recall the coordinate and null-tetrad system used in \cite{BST} and prove Theorems \ref{one} and \ref{two}, to show that, in this setting, periodic solutions are actually stationary. The proof is much as in \cite{BST}: one shows inductively that all radial derivatives of all metric components at $\scri^-$ are $v$-independent so that $K=\partial/\partial v$ is a Killing vector. In section 6, we discuss inheritance and prove Theorem \ref{three} to show that stationarity is necessarily inherited in an analytic, weakly-asymptotically-flat, Einstein-massless-KG solution.

In order to be able to follow clearly the arguments in the main text, in Appendix A we review all Newman-Penrose equations for a general source: that is, the commutation relations of the NP operators, and the Ricci and Bianchi identities. In Appendix B the conformally rescaled scalar wave equations and conformal Bianchi identities for the massless scalar field are written down in unphysical space in manifestly regular form. The regular conformal Bianchi identities for conformally invariant scalar fields follow in fact from the conformal Bianchi identities for any matter field for which the Ricci spinor behaves as ${\cal O}(\Omega^2)$ at $\scri$. The projections of the Bianchi identities are given in section B4 of Appendix B. The asymptotic form of the solutions of the Einstein-massless-scalar field equations at future null infinity $\scri^+$ is discussed in Appendix C; this is used in section 4 in the derivation of the Bondi mass and the mass-loss formula for both massless-Klein-Gordon field and conformal-scalar field. Finally, in Appendix D we give some examples of exact solutions of the Einstein-conformal-scalar equations and discuss the  possible presence of singularities.

\section{The massless KG field}

\subsection{Basic relations}

First we investigate the complex scalar field which satisfies the massless Klein-Gordon equation in the physical spacetime,

\begin{eqnarray}
\tilde{\Box}\,\tilde{\phi} &=& 0.\label{KGEqPhys}
\end{eqnarray}
We shall consistently use the tilde to indicate quantities in the physical space-time, untilded quantities referring to the rescaled, unphysical space-time. The energy-momentum tensor whose conservation is implied by this equation is

\begin{eqnarray}
\widetilde{T}_{ab} &=& \frac{1}{4\pi}\;\left[ 2\;\left(\widetilde{\nabla}_{(a}\tilde{\phi}\,\right)\left(\widetilde{\nabla}_{b)}\tilde{\bar{\phi}}\right)\;-
\;\tilde{g}_{ab}\tilde{g}^{cd}\,\left(\widetilde{\nabla}_{c}\tilde{\phi}\right)\left(\widetilde{\nabla}_d\tilde{\bar{\phi}}\right)\,\right].\label{KG-Tab}
\end{eqnarray}

First we have to determine the conformal behaviour of the scalar field. Since the wave equation (\ref{KGEqPhys}) is not conformally invariant, there is a priori no preferred choice. However, since near $\scri^-$ ($\tilde{r}\rightarrow\infty$) the radiative part of the field behaves as

\begin{eqnarray}
\tilde{\phi} &\sim&\frac{1}{\tilde{r}},
\end{eqnarray}
and we wish to have a non-vanishing regular unphysical field on $\scri^-$, we define

\begin{eqnarray}
\tilde{\phi} &=& \Omega\;\phi.\label{PhiConfTransf}
\end{eqnarray}
In the following, we employ the notation

\begin{eqnarray}
\tilde{\fii}_{AA^\prime}\;=\;\nabla_{AA^\prime}\tilde{\phi}, \;\;\;\;\fii_{AA^\prime}\;=\;\nabla_{AA^\prime}\phi, \;\;\;\;s_{AA^\prime}=\nabla_{AA^\prime}\Omega,\label{DefSa}
\end{eqnarray}
and using the NP formalism \footnote{The explicit expressions for the NP tetrad, the corresponding spin basis, the NP operators, etc., in the coordinate system $(v,r,\theta,\phi)$ used in the following are introduced in section 3 and Appendix A of paper I and repeated in section 5 below.} we denote the components of the fields $s_a$ and $\fii_a$ by special symbols:

\begin{eqnarray}
\fl
\begin{array}{llll}
D\Omega=s_{00^\prime}=S_0,&\delta\Omega=s_{01^\prime}=S_1,&\bar{\delta}\Omega=s_{10^\prime}=S_{\bar{1}}=\bar{S_1},&\Delta\Omega=s_{11^\prime}=S_2,\\
D\phi=\fii_{00^\prime}=\fii_0,&\delta\phi=\fii_{01^\prime}=\fii_1,&\bar{\delta}\phi=\fii_{10^\prime}=\fii_{\bar{1}},&\bar{\Delta}\phi=\fii_{11^\prime}=\fii_2,\\
\end{array}
\end{eqnarray}
and correspondingly with tildes in the physical spacetime.

\noindent In this notation, the spinor form of Einstein's equations in the physical spacetime is

\begin{eqnarray}
\eqalign{
\tilde{\Phi}_{ABA^\prime B^\prime} &=\;2\,\tilde{\fii}_{(A(A^\prime}\,\tilde{\bar{\fii}}_{B^\prime)B)},\\
6\,\tilde{\Lambda} &=\;-\,\tilde{\fii}_c\,\tilde{\bar{\fii}}^c,\label{EinstEqs1}}
\end{eqnarray}
the components of the Ricci spinor with respect to the spin basis are

\begin{eqnarray}
\eqalign{
\tilde{\Phi}_{00} &=\, 2\,\tilde{\fii}_0 \, \tilde{\bar{\fii}}_0,\\
\tilde{\Phi}_{01} &=\, 2\,\tilde{\fii}_{(0}\,\tilde{\bar{\fii}}_{1)},\\
\tilde{\Phi}_{02} &=\, 2\,\tilde{\fii}_{1}\,\tilde{\bar{\fii}}_{1},\\
\tilde{\Phi}_{11} &=\, \tilde{\fii}_{(0}\,\tilde{\bar{\fii}}_{2)}\;+\;\tilde{\fii}_{(1}\,\tilde{\bar{\fii}}_{\bar{1})},\\
\tilde{\Phi}_{12} &=\, 2\,\tilde{\fii}_{(1}\,\tilde{\bar{\fii}}_{2)},\\
\tilde{\Phi}_{22} &=\, 2\,\tilde{\fii}_{2}\,\tilde{\bar{\fii}}_{2},\\
\label{RicciComps1}}
\end{eqnarray}
and the scalar curvature is

\begin{eqnarray}
\tilde{\Lambda} &= \,\frac{1}{3}\,\left[ -\,\tilde{\fii}_{(0} \,\tilde{\bar{\fii}}_{2)} \;+\; \tilde{\fii}_{(1} \,\tilde{\bar{\fii}}_{\bar{1})}\right].\label{Lambda1}
\end{eqnarray}

\subsection{The conformal Einstein-massless-KG equations}\label{ssCEKG}
In this subsection, we find a system of equations regular at $\scri$ for all unphysical quantities. This system, by
analogy with Friedrich's `conformal Einstein equations' \cite{F} we shall call the `conformal Einstein-massless-KG equations'.

In \cite{BST} we derived the physical Bianchi identities expressed in terms of the unphysical quantities as

\begin{eqnarray}
\fl
\Omega^2\,\nabla^D_{A^\prime} \psi_{ABCD} &=\;\Omega\,\nabla^{B^\prime}_{(C}\,\Phi_{AB)A^\prime B^\prime}\;+\;s^{B^\prime}_{(C}\,\Phi_{AB)A^\prime B^\prime}\;+\;\nabla^{B^\prime}_{(C}\,\nabla_{A(A^\prime}\,s_{B^\prime)B)}.
\label{ConfBI1}
\end{eqnarray}

\noindent where $s_{AA^\prime}$ is given by (\ref{DefSa}), $\Phi_{ABA^\prime B^\prime}$ is the Ricci spinor and $\psi_{ABCD}=\Omega^{-1}\Psi_{ABCD}$ is the rescaled Weyl spinor (see eq. (6) in I). Using the rule for the conformal transformation of the Ricci spinor,

\begin{eqnarray}
\nabla_{A(A^\prime}s_{B^\prime)B} &=\;\Omega\,\tilde{\Phi}_{ABA^\prime B^\prime}\;-\;\Omega\,\Phi_{ABA^\prime B^\prime},\label{ConfTransRicci}
\end{eqnarray}

\noindent we find

\begin{eqnarray}
\nabla^D_{A^\prime}\,\psi_{ABCD} &=\;\Omega^{-2}\,s^{B^\prime}_{(C}\,\tilde{\Phi}_{AB)A^\prime B^\prime}\;+\;\Omega^{-1}\,\nabla^{B^\prime}_{(C}\,\tilde{\Phi}_{AB)A^\prime B^\prime}.\label{ConfBI2}
\end{eqnarray}

\noindent The right hand side of this equation is not manifestly regular on $\scri$, while the left hand side is regular by assumption of asymptotic flatness.

Next we express the physical Ricci spinor via the unphysical quantities,

\begin{eqnarray}
\eqalign{
\tilde{\Phi}_{ABA^\prime B^\prime} &=\;2\,\Omega^2\,\fii_{(A(A^\prime}\bar{\fii}_{B^\prime)B)}\;+\; 2\,\phi\,\bar{\phi}\,s_{(A(A^\prime}\,s_{B^\prime)B)}\\
&+\;2\,\Omega\,\bar{\phi}\,\fii_{(A(A^\prime}\,s_{B^\prime)B)}\;+\;2\,\Omega\,\phi\,\bar{\fii}_{(A(A^\prime}\,s_{B^\prime)B)}.
\label{PhysRicSp1}
}
\end{eqnarray}

\noindent and insert this expression into (\ref{ConfBI2}).

In order to simplify the resulting equations, we introduce following notation: let $X_a, Y_a$ and $Z_a$ be arbitrary vector fields and define

\begin{eqnarray}
\SYMM{XYZ} &=\;X^{B^\prime}_{(C}\,Y_{A(A^\prime}\,Z_{B^\prime)B)}.\label{SYMsymbol}
\end{eqnarray}

\noindent The expression $\SYMM{XYZ}$ is obviously symmetric in $YZ$. It is straightforward to derive the relation

\begin{eqnarray}
\SYMM{XYZ}\;+\;\SYMM{ZXY}\;+\;\SYMM{YZX} &=\;0,\label{SYMMcycl}
\end{eqnarray}

\noindent with special case $\SYMM{XXX}=0$.

After inserting the Ricci spinor (\ref{PhysRicSp1}) into the Bianchi identities (\ref{ConfBI2}), we arrive at

\begin{eqnarray}
\fl
\eqalign{
\nabla^D_{A^\prime}\,\psi_{ABCD} &=\;2\,\Omega^{-1}\,\left( 2\,\bar{\phi}\,\SYMM{s\fii s} \;+\;2\,\phi\,\SYMM{s\bar{\fii}s} \;+\;\phi\,\SYMM{\bar{\fii} s\, s} \;+\;\bar{\phi}\,\SYMM{\fii s\,s}\;+\;\phi\,\bar{\phi}\,\SYMM{\nabla s \, s} \right) \\
&+\;6\,\SYMM{s\fii\bar{\fii}}\;+\;2\,\SYMM{\bar{\fii}\fii\,s}\;+\;2\,\SYMM{\fii\bar{\fii}\,s} \;+\;
2\,\bar{\phi}\,\SYMM{\nabla\fii\,s}\;+\;2\,\phi\,\SYMM{\nabla\bar{\fii}\,s}\\
&+\;2\,\Omega\,\SYMM{\nabla\fii\,\bar{\fii}}.
}
\end{eqnarray}

\noindent This can be simplified using identity (\ref{SYMMcycl}):

\begin{eqnarray}
\eqalign{
\nabla^D_{A^\prime}\psi_{ABCD} &=\;2\,\phi\,\bar{\phi}\,\Omega^{-1}\,\SYMM{\nabla\,s\,s} \;+\;2\,\Omega\,\SYMM{\nabla\,\fii\,\bar{\fii}}\\
&+\;4\,\SYMM{s\,\fii\,\bar{\fii}}\;+\;2\,\bar{\phi}\,\SYMM{\nabla\,\fii\,s}\;+\;2\,\phi\,\SYMM{\nabla\,\bar{\fii}\,s},
}
\end{eqnarray}

\noindent where, e.g. $\SYMM{\nabla\,\fii\, s} = \nabla^{B^\prime}_{(C}\left( \fii_{A(A^\prime}\, s_{B^\prime)B)}\right)$, so $\nabla$ acts on both $\fii$ and $s$. The last equation is still formally singular on $\scri^-$ because of the factor $\Omega^{-1}$, but using (\ref{ConfTransRicci}) we finally obtain

\begin{eqnarray}
\fl
\eqalign{
\nabla^D_{A^\prime}\psi_{ABCD} &=\; 2\,\phi\,\bar{\phi}\,s^{B^\prime}_{(C}\Phi_{AB)A^\prime B^\prime}\;+\;4\,\SYMM{s\,\fii\,\bar{\fii}}\;+\;2\,\phi\,\SYMM{\nabla\,s\,\bar{\fii}} \;+\;2\,\bar{\phi}\,\SYMM{\nabla\,s\,\fii}\\
&+\;4\,\Omega\,\left[ \frac{1}{2}\,\SYMM{\nabla\,\fii\,\bar{\fii}}\;-\;\phi\,\bar{\phi}^2\,\SYMM{s\,\fii\,s}\;-\;\bar{\phi}\,\phi^2\,\SYMM{s\,\bar{\fii}\,s}\right]\\
&-\;4\,\Omega^2\,\phi\,\bar{\phi}\,\SYMM{s\,\fii\,\bar{\fii}},\label{ConfBI0}}
\end{eqnarray}
which is manifestly smooth at $\scri$.

Next we wish to derive equations for the conformal factor. The commutator of covariant derivatives annihilates scalars, so contracting $\nabla_{[a}\nabla_{b]}\Omega=0$ with $\epsilon^{A^\prime B^\prime}$ gives the relation

\begin{eqnarray}
\nabla_{A^\prime(A}\,s^{A^\prime}_{B)} &=& 0.
\end{eqnarray}

\noindent By decomposing $\nabla_{AA^\prime}s_{BB^\prime}$ into its symmetric and antisymmetric parts and using the equation above, we obtain

\begin{eqnarray}
\nabla_{AA^\prime} \,s_{BB^\prime} &=& \nabla_{(A(A^\prime}\,s_{B^\prime)B)}\;\;+\;\;\frac{1}{4}\,\epsilon_{AB}\,\epsilon_{A^\prime B^\prime}\,\Box\Omega.
\end{eqnarray}

\noindent The first term on the r.h.s. is given by (\ref{ConfTransRicci}). We now define the quantity (cf. eq. (15) in I)

\begin{eqnarray}
F &=& \frac{1}{2}\,\Omega^{-1}\,g^{ab}\,s_a\,s_b,\label{DefOfF}
\end{eqnarray}

\noindent which is regular on $\scri$. The rule for the conformal transformation of the scalar curvature can be written in the form (eq. (16) in I)

\begin{eqnarray}
\Box\Omega &=& 4\,\Omega\,\Lambda\;-\;4\,\Omega^{-1}\,\tilde{\Lambda}\;+\;4\,F.
\end{eqnarray}

\noindent We thus have found an expression for the second derivatives of the conformal factor $\Omega$:

\begin{eqnarray}\fl
\nabla_{AA^\prime}s_{BB^\prime} \;=\; \Omega\,\tilde{\Phi}_{ABA^\prime B^\prime}\;-\;\Omega\,\Phi_{ABA^\prime B^\prime}\;+\;\epsilon_{A B}\,\epsilon_{A^\prime B^\prime}\,\left(\Omega\Lambda-\Omega^{-1}\tilde{\Lambda}+F\right).\label{DDs}
\end{eqnarray}

The last expression contains a term $\Omega^{-1}\tilde{\Lambda}$ which again seems to be singular on $\scri$. This is not the case, however, since by (\ref{EinstEqs1}), (\ref{PhiConfTransf}) and (\ref{DefOfF}) we have

\begin{eqnarray}
\tilde{\Lambda} &=\;-\,\frac{1}{6}\,\Omega^3\,\left[ \Omega\,\fii_c\,\bar{\fii}^c +  \phi\, \bar{\fii}_c\, s^c + \bar{\phi}\, \fii_c \,s^c + 2 \,\phi\, \bar{\phi}\,F\right].\label{LambdaOrd}
\end{eqnarray}

\noindent The physical scalar curvature is therefore manifestly at least $\ord{\Omega^3}$.

The projections of the last equation are written down explicitly in Appendix B, eqs. (\ref{s01})--(\ref{s23}). Now we wish to derive equations governing the quantity $F$. The contracted Ricci identities read

\begin{eqnarray}
\nabla_{a}\,\Box \Omega\;\;-\;\;\nabla_b\nabla_a\,s^b &=& R_a^d\,s_d.
\end{eqnarray}

\noindent Using the spinor decomposition of the Ricci tensor

\begin{eqnarray}
R_{ab}&=\;-\,2\,\Phi_{A B A^\prime B^\prime}\;+\;6\,\Lambda\,\epsilon_{AB}\,\epsilon_{A^\prime B^\prime}.
\end{eqnarray}

\noindent and the expression (\ref{DDs}), we find after some arrangements

\begin{eqnarray}
\fl
\eqalign{
\nabla_{AA^\prime}F &=\; \frac{1}{3}\,\Omega^{1}\,\nabla^{BB^\prime}\,\tilde{\Phi}_{ABA^\prime B^\prime}\;+\;\frac{1}{3}\,s^{BB^\prime}\,\tilde{\Phi}_{ABA^\prime B^\prime}\;-\;s^{BB^\prime}\,\Phi_{ABA^\prime B^\prime}\\
&+\;\Lambda\,s_{AA^\prime}\;-\;\Omega^{-2}\,\tilde{\Lambda}\,s_{AA^\prime}\;+\;\Omega^{-1}\,\nabla_{AA^\prime}\tilde{\Lambda}.}
\end{eqnarray}

\noindent The first term on the r.h.s. can be rewritten as

\begin{eqnarray}
\fl
\nabla^{BB^\prime}\tilde{\Phi}_{ABA^\prime B^\prime} &=\;\Omega^{-2}\,\tilde{\nabla}^{BB^\prime}\tilde{\Phi}_{ABA^\prime B^\prime}\;+\;2\,\Omega^{-1}\,\tilde{\Phi}_{ABA^\prime B^\prime}\,s^{BB^\prime}.
\end{eqnarray}

\noindent Now we employ the contracted physical Bianchi identities $\tilde{\nabla}^{AA^\prime}\tilde{\Phi}_{ABA^\prime B^\prime} = - 3\tilde{\nabla}_{BB^\prime}\tilde{\Lambda}$ and obtain

\begin{eqnarray}
\fl
\nabla_{AA^\prime} F \;=\; s^{BB^\prime}\,\tilde{\Phi}_{ABA^\prime B^\prime} \;-\;s^{BB^\prime}\,\Phi_{ABA^\prime B^\prime}\;+\;(\Lambda-\Omega^{-2}\,\tilde{\Lambda})\,s_{AA^\prime}.\nonumber\\\label{FDers}
\end{eqnarray}

\noindent The projections of this equation can be found in Appendix B, eqs. (\ref{F1})--(\ref{F3}).

Finally we derive conformal equations for the field $\fii_{AA^\prime}$. The expression $(\Box + 4 \Lambda) \phi$ is conformally invariant with the conformal weight $3$, so

\begin{eqnarray}
\Box\phi &=\;-\,4\,(\Lambda - \Omega^{-2}\,\tilde{\Lambda})\,\phi,
\end{eqnarray}

\noindent where we used the wave equation (\ref{KGEqPhys}) in the physical spacetime. The symmetric part of $\nabla_A^{A^\prime}\fii_{BA^\prime}$ is zero and we find

\begin{eqnarray}
\nabla_A^{A^\prime}\fii_{BA^\prime} &=\;-\,\frac{1}{2}\,\epsilon_{AB}\,\Box\phi.
\end{eqnarray}

\noindent Combining the last two equations we arrive at

\begin{eqnarray}
\nabla_A^{A^\prime}\fii_{BA^\prime} &=\;2\,\left(\Lambda-\Omega^{-2}\tilde{\Lambda}\right)\,\phi\,\epsilon_{AB}.\label{ConfEqScField}
\end{eqnarray}

To summarise: in the unphysical spacetime we have the following variables: $\left\{\Omega, \phi, s_a, \fii_a, F, \psi_{ABCD}, \Phi_{ABA^\prime B^\prime}, \Lambda\right\}$. The evolution of these quantities is given by eqs. (\ref{ConfBI0}), (\ref{DDs}),  (\ref{FDers}) and (\ref{ConfEqScField}), together with the contracted Bianchi identities

\begin{eqnarray}
\nabla^{AA^\prime}\Phi_{ABA^\prime B^\prime} &=\;-\,3\,\nabla_{BB^\prime}\Lambda.
\end{eqnarray}

\subsection{Potentials}\label{ssPot}

The massless KG equation (\ref{KGEqPhys}) can be generalized to include self-interactions of the scalar field by adding a potential term to the energy-momentum tensor (\ref{KG-Tab}),

\[
\widetilde{T}_{ab} \;\mapsto\;\widetilde{T}_{ab}\;+\;\frac{1}{4\pi}\,\tilde{g}_{ab}\,V(\tilde{\phi}, \tilde{\bar{\phi}}),
\]

\noindent so that the field equation acquires the form

\[
\widetilde{\Box}\tilde{\phi}\;+\;\frac{\partial V}{\partial \tilde{\bar{\phi}}}\;=\;0.
\]

Since the potential term in the energy-momentum tensor is proportional to the metric, it will contribute to the scalar curvature $\tilde{\Lambda}$, but not to the trace-free Ricci spinor. The new form of Einstein's equations (\ref{EinstEqs1}) is therefore

\begin{eqnarray}
\eqalign{
\tilde{\Phi}_{ABA^\prime B^\prime}&=\;2\,\tilde{\fii}_{(A(A^\prime}\,\tilde{\bar{\fii}}_{B^\prime)B)},\\
6\,\tilde{\Lambda} &=\;- \tilde{\fii}_c\,\tilde{\bar{\fii}}{}^c\;+\;2\,V.
\label{EinstEqsV}
}
\end{eqnarray}

For our proof we require $\phi=\Omega^{-1}\tilde{\phi}$ and $\Omega^{-3}\tilde{\Lambda}$ to be regular on $\scri^-$, cf. (\ref{PhiConfTransf}), (\ref{FDers}) and (\ref{LambdaOrd}). From (\ref{EinstEqsV}) we can see that this will be satisfied, if $\Omega^{-3}V$ is regular on $\scri^-$. In this case the proof works without change. An example is massless $\phi^4-$theory, where $V=(\tilde{\phi}\tilde{\bar{\phi}})^2 = \ord{\Omega^4}$.

If there is a mass term $m^2 \tilde{\phi} \tilde{\bar{\phi}}$ in $V$, the asymptotic behaviour of the unphysical field changes to

\[
\phi\;=\;\ord{e^{-m\tilde{r}}},
\]

\noindent so

\[
\Omega^{-3}\, m^2\,\tilde{\phi}\,\tilde{\bar{\phi}}\;\sim\;m^2\,\tilde{r}\,e^{-2m\tilde{r}},
\]

\noindent which is regular. The field $\phi$ is now not analytic at $\scri^-$, so our argument does not apply to this case, which is the class including the boson stars of \cite{BW}. Notice, however, that in general the asymptotic behaviour of massive fields at $\scri$ is a subtle question which appears to be carefully analyzed only at the level of linearized theory \cite{Winicour}.

\section{The conformal-scalar field}

\subsection{Basic relations}

Consider now the conformal-scalar field, by which we mean a scalar field satisfying the equation

\begin{eqnarray}
\left(\widetilde{\Box}\;+\;\frac{1}{6}\,\tilde{R}\,\right)\tilde{\phi}\;\;\equiv\;\;\left( \widetilde{\Box}\;+\;4\,\tilde{\Lambda} \right)\tilde{\phi} &=& 0.\label{ConfInvWaveEq}
\end{eqnarray}

\noindent This is conformally invariant if $\phi$ transforms as (\ref{PhiConfTransf}), i.e. $\tilde{\phi}=\Omega\,\phi$. For simplicity we assume the field $\phi$ to be real, but the procedure is easily generalized to complex $\phi$. The energy-momentum tensor conserved due to equation (\ref{ConfInvWaveEq}) is (see \cite{NP}, \cite{CCJ} or \cite{PR}, Volume II, page 125)

\begin{eqnarray}
\widetilde{T}_{ab} &=\;\frac{1}{4\pi}\,\left[ 2\,\tilde{\fii}_{A(A^\prime}\,\tilde{\fii}_{B^\prime)B}\;-\;\tilde{\phi}\,\widetilde{\nabla}_{A(A^\prime}\,\tilde{\fii}_{B^\prime)B}\;+\;\tilde{\phi}{}^2\,\tilde{\Phi}_{ABA^\prime B^\prime}\right].
\label{ConfCoupledTab}
\end{eqnarray}

\noindent Furthermore, this energy momentum tensor also has good conformal behavior, rescaling as
\[\widetilde{T}_{ab}=\Omega^2T_{ab},\]
but it will not satisfy any of the usual energy conditions. We shall return to this point. We take Einstein's equations to be, as usual,
\[\tilde{\Phi}_{ab}+3\tilde{\Lambda}\tilde{g}_{ab}=4\pi \tilde{T}_{ab},\]
then we can solve to find
\begin{eqnarray}
\eqalign{
\tilde{\Phi}_{ABA^\prime B^\prime} &=\;\left(1-\tilde{\phi}{}^2\right)^{-1}\left[ 2\,\tilde{\fii}_{(A(A^\prime}\,\tilde{\fii}_{B^\prime)B)}\;-\;\tilde{\phi}\,\tilde{\nabla}_{(A(A^\prime}\,\tilde{\fii}_{B^\prime)B)}\right],\\
\tilde{\Lambda} &=\;0.
\label{EinstEQS2}}
\end{eqnarray}
These equations are singular when $\tilde{\phi}^2=1$ but there are known solutions which avoid this singularity \cite{bek1}, \cite{bek2} (for explicit examples, see Appendix D) and it is known that there is a well-posed initial value problem \cite{hub} which with suitable data extends to $\scri^+$.  We shall therefore assume that we have an asymptotically-flat solution, periodic in time, with $\tilde{\phi}$ tending to zero at infinity, so that $\tilde{\phi}^2<1$ everywhere.

In the absence of the Dominant Energy Condition,
it isn't clear that any version of the Positive Mass Theorem holds but there is some reason to expect a positive global energy. To see this, integrate the energy density over an asymptotically-flat maximal space-like hypersurface $\Sigma$ (assuming for the moment that one exists)
 with normal $N^a$. Note from (\ref{ConfCoupledTab}) and (\ref{EinstEQS2}) that
\begin{eqnarray}
\widetilde{T}_{ab} =\;\frac{1}{4\pi}\,\frac{1}{1-\tilde{\phi}^2}\left[ 2\,\tilde{\fii}_{A(A^\prime}\,\tilde{\fii}_{B^\prime)B}\;-\;\tilde{\phi}\,\widetilde{\nabla}_{A(A^\prime}\,\tilde{\fii}_{B^\prime)B}\right].
\label{TabConfInv}
\end{eqnarray}
\noindent Then a measure of total energy at $\Sigma$ is
\begin{eqnarray}
\fl
\eqalign{
E:&=\int \widetilde{T}_{ab}N^aN^b\d\Sigma
\\&=\frac{1}{4\pi}\int \frac{1}{1-\tilde{\phi}{}^2}
\left[ (N^a\tilde{\fii}_a)^2-\frac{1}{2}\tilde{g}^{ab}\tilde{\fii}_a\tilde{\fii}_b-
\tilde{\phi}\,N^a\,N^b\,\widetilde{\nabla}_a\,\tilde{\fii}_b \right]\d\Sigma.
\label{EE}
}
\end{eqnarray}
\noindent Now,
\[\fl
N^a\,N^b\,\widetilde{\nabla}_a\tilde{\fii}_b \;=\; (h^{ab}\;+\;\tilde{g}^{ab})\widetilde{\nabla}_a\tilde{\fii}_b\;=\;h^{ij}\,\widetilde{\nabla}_i\,\tilde{\fii}_j\;=\;h^{ij}D_i\tilde{\fii}_j\;+\;(N^a\tilde{\fii}_a)K,
\]

\noindent where $D_i$ is the derivative operator associated with the three-dimensional metric $h_{ij}$ induced on $\Sigma$, $K$ is the trace of the extrinsic curvature which vanishes for a maximal surface, and we have used $\widetilde{\Box}\tilde{\phi}=0$.

\noindent Note also

\[
\tilde{g}^{ab}\tilde{\fii}_a\tilde{\fii}_b
=(N^a \tilde{\fii}_a)^2-h^{ij}\tilde{\fii}_i\tilde{\fii}_j,
\]
\noindent and integrate by parts in (\ref{EE}) to find
\[
E\;=\;\frac{1}{4\pi}\int \d \Sigma\; (1-\tilde{\phi}^2)^{-1}\left[ \frac{3}{2}(N^a\tilde{\fii}_a)^2 \;+\;\frac{1}{2}\,\frac{3+\tilde{\phi}{}^2}{1-\tilde{\phi}{}^2}\,h^{ij}\tilde{\fii}_i\tilde{\fii}_j\right],
\]
which is manifestly non-negative. Thus on a maximal surface the global energy is positive without local positivity. Positive energy also holds for hyperplanes in Minkowski space with $T_{ab}$ as in (\ref{ConfCoupledTab}) and we shall see something similar below, namely that, while the Bondi mass-loss is not necessarily positive at any particular cut, nonetheless the mass-loss integrated over a period in a periodic space-time is non-negative.

The Ricci spinor written in terms of unphysical quantities reads

\begin{eqnarray}
\fl
\eqalign{
\left(1 - \Omega^2 \phi^2\right)\tilde{\Phi}_{ABA^\prime B^\prime} &=\; 2\,\Omega^2\,\fii_{(A(A^\prime}\,\fii_{B^\prime)B)}\;-\;\Omega^2\,\phi\,\nabla_{(A(A^\prime}\,\fii_{B^\prime)B)}\\
&-\;\Omega\,\phi^2\,\nabla_{(A(A^\prime}\,s_{B^\prime)B)}.\label{PhysRicSp}
}
\end{eqnarray}

\noindent Let us define the ``rescaled Ricci spinor"

\begin{eqnarray}
\phi_{ABA^\prime B^\prime} &=\;\Omega^{-2}\,\tilde{\Phi}_{ABA^\prime B^\prime},\label{RescaledRicci}
\end{eqnarray}
which should be distinguished from the unphysical Ricci spinor. Substituting the rule for conformal transformation of the Ricci spinor (\ref{ConfTransRicci}) into (\ref{PhysRicSp}) we arrive at the following simple expression for the rescaled Ricci spinor:

\begin{eqnarray}
\phi_{ABA^\prime B^\prime} &=& 2\,\fii_{(A(A^\prime}\,\fii_{B^\prime)B)}\;-\;\phi\,\nabla_{(A(A^\prime}\fii_{B^\prime)B)}\;+\;\phi^2\,\Phi_{ABA^\prime B^\prime}.\nonumber\\\label{RescRicciSp}
\end{eqnarray}

\noindent This spinor is regular on $\scri^-$. Notice that we do not write the tilde over $\phi_{ABA^\prime B^\prime}$ (as we wrote over $\tilde{\Phi}_{ABA^\prime B^\prime}$ in (\ref{EinstEqs1})), since we expect that the physical Ricci spinor has been already expressed in terms of the unphysical quantities and the following relations become simpler. The components of $\phi_{ABA^\prime B^\prime}$ with respect to the spin basis are

\begin{eqnarray}
\fl
\phi_{00} =\;2\fii_0^2-\phi\left[D\fii_0 - (\eps+\bar{\eps})\fii_0 + \bar{\kappa}\fii_1 + \kappa\fii_{\b1}\right]+\phi^2\,\Phi_{00},\nonumber\\
\fl
\phi_{01} =\;2\fii_0\fii_1 - \frac{1}{2}\phi\left[ D\fii_1 +\delta\fii_0 - (\bar{\alpha}+\beta+\bar{\pi})\fii_0 + \kappa \fii_2 + (\bar{\rho}-\eps+\bar{\eps})\fii_1 + \sigma \fii_{\b1} \right]+\phi^2\,\Phi_{01},\nonumber\\
\fl
\phi_{02} =\;2\fii_1^2 - \phi\left[\delta\fii_1-\bar{\lambda}\fii_0 + \sigma\fii_2 + (\bar{\alpha}-\beta)\fii_1\right]+\phi^2\,\Phi_{02},\nonumber\\
\fl
\phi_{12} =\;2\fii_1\fii_2 - \frac{1}{2}\phi\left[\Delta\fii_1 + \delta\fii_2 - \bar{\nu}\fii_0 + (\beta + \tau + \bar{\alpha})\fii_2 + (\bar{\gamma}-\gamma-\mu)\fii_1 - \bar{\lambda}\fii_{\b1}\right]+\phi^2\,\Phi_{12},\nonumber\\
\fl
\phi_{22} =\;2\fii_2^2 - \phi\left[\Delta\fii_2 + (\gamma+\bar{\gamma})\fii_2 - \nu \fii_1 - \bar{\nu}\fii_{\b1}\right]+\phi^2\,\Phi_{22},\\
\fl
\phi_{11} =\;\fii_0\fii_2 + \fii_1 \fii_{\b1} - \frac{1}{4}\phi\left[D \fii_2 +\Delta\fii_0 + \delta\fii_{\b1} + \bar{\delta}\fii_1 - (\gamma+\bar{\gamma}+\mu+\bar{\mu})\fii_0\right]\nonumber \\
\fl\;\;\;-\;\frac{1}{4}\phi\left[(\rho+\bar{\rho}+\eps+\bar{\eps})\fii_2 + (\bar{\tau}-\alpha+\bar{\beta}-\pi)\fii_1 + (\tau - \bar{\alpha} + \beta - \bar{\pi})\fii_{\b1}\right]+\phi^2\,\Phi_{11}.\nonumber
\end{eqnarray}

\subsection{The conformal Einstein-conformal-scalar equations}
Now, as in subsection \ref{ssCEKG}, we obtain a system of conformal Einstein equations, regular in the unphysical, rescaled space-time and equivalent to the Einstein-conformal-scalar equations.

In order to derive the conformal Bianchi identities for the conformal-scalar field we return to the general physical Bianchi identities (\ref{ConfBI2}). Using the rescaled Ricci spinor instead of $\tilde{\Phi}_{ABA^\prime B^\prime}$ the Bianchi identities become

\begin{eqnarray}
\nabla_{A^\prime}^D\,\psi_{ABCD} &=& 3\,s^{B^\prime}_{(C}\,\phi_{AB)A^\prime B^\prime}\;+\;\Omega\,\nabla^{B^\prime}_{(C}\phi_{AB)A^\prime B^\prime}.\label{ConfCoupledBI}
\end{eqnarray}

\noindent Projections of these equations on the spin basis can be found in Appendix B, eqs. (\ref{ConfCpldBI1})--(\ref{ConfCpldBI8}).

Next we turn to the contracted Bianchi identities in the physical spacetime

\begin{eqnarray}
\tilde{\nabla}^{BB^\prime}\tilde{\Phi}_{ABA^\prime B^\prime} &=\; -\;3\,\tilde{\nabla}_{AA^\prime}\,\tilde{\Lambda},
\end{eqnarray}

\noindent where $\tilde{\Lambda}=0$ by (\ref{EinstEQS2}). Following the rules for conformal transformation of the covariant derivative we find that the left hand side transforms like

\begin{eqnarray}
\tilde{\nabla}^{BB^\prime}\tilde{\Phi}_{ABA^\prime B^\prime} &=& \Omega^2\,\nabla^{BB^\prime}\tilde{\Phi}_{ABA^\prime B^\prime}\;-\;2\,\Omega\,s^{BB^\prime}\,\tilde{\Phi}_{ABA^\prime B^\prime}
\end{eqnarray}

\noindent or, using (\ref{RescaledRicci}),

\begin{eqnarray}
\tilde{\nabla}^{BB^\prime}\tilde{\Phi}_{ABA^\prime B^\prime} &=& \Omega^4\,\nabla^{BB^\prime}\phi_{ABA^\prime B^\prime},
\end{eqnarray}

\noindent and thus the contracted Bianchi identities have a simple form, just as in the physical spacetime:

\begin{eqnarray}
\nabla^{BB^\prime}\phi_{ABA^\prime B^\prime} &=& 0. \label{ConfCpldContrBianchi}
\end{eqnarray}

\noindent Projections of these equations on the spin basis can be obtained from the Bianchi identities (\ref{BIC1})--(\ref{BIC3}) by deleting terms containing $\Lambda$ and replacing $\Phi_{mn}\mapsto \phi_{mn}$.


\section{Bondi mass}

One of the necessary ingredients in this work is to find restrictions which the assumption of periodicity imposes on the Bondi mass. As long as the Bondi mass $M_B(u)$ on $\scri^+$ is a non-increasing function of retarded time $u$, it can be periodic only if it is constant. In \cite{BST} we used the well-known formula for the Bondi mass of an electro-vacuum spacetime\footnote{In fact, in \cite{BST} we constructed the proof -- and the same will be done here -- at $\scri^-$ where the Bondi mass is non-decreasing but it is straightforward to get one from the other. Since it is more common to work at $\scri^+$, in this section we discuss the Bondi mass there.},

\begin{eqnarray}
M_B(u) &=  -\;\frac{1}{2\sqrt{\pi}}\oint\d S\;\left( \Psi_2^0\;+\;\sigma^0\,\dot{\bar{\sigma}}{}^0\right),
\end{eqnarray}

\noindent when its time decrease is given by the ``mass-loss" formula

\begin{eqnarray}
\dot{M}_B(u) &=  -\;\frac{1}{2\sqrt{\pi}}\oint\d S\;\left( \dot{\sigma}^0\,\dot{\bar{\sigma}}{}^0\;+\;\phi_2^0\,\bar{\phi}_2^0\right).
\end{eqnarray}

\noindent This expression is manifestly non-positive. To achieve periodicity of the Bondi mass we thus had to set $\dot{\sigma}^0=0$ and $\phi_2^0=0$. The loss of Bondi mass due to the gravitational radiation is described by the news function $-\dot{\bar{\sigma}}^0$ and the electromagnetic contribution by the quantity $\phi_2^0$. Periodicity thus requires the absence of both gravitational and electromagnetic radiation.

In order to repeat this reasoning in the case of spacetimes with scalar fields, we need the appropriate formula for Bondi mass-loss. The gravitational contribution will again be expressed by the news function and there will be a contribution from the matter. The energy flux due to the matter is described by the energy-momentum tensor $T_{ab}$ (omitting tildes for clarity, in this subsection only). If we write, using the NP formalism,

\begin{eqnarray}
T_{ab}&=\;A \,l_{a}\,l_{b}\;+\;B\,n_{(a}l_{b)} \;+\;C\,n_{a}\,n_{b}\;+\;\cdots,
\end{eqnarray}

\noindent then the component $A = T_{ab} n^a n^b$ is the energy radiated out of $\scri^+$ (recall that $n^a$ is tangential to $\scri^+$, $l^a$ points into the space-time towards $\scri^-$). In terms of the Ricci spinor NP component we get

\begin{eqnarray}
T_{ab} \,n^a\,n^b &\propto\;\Phi_{22}.
\end{eqnarray}

\noindent For the complex scalar field $\phi$, $\Phi_{22} \propto \dot{\phi} \,\bar{\dot{\phi}}$, where dot means the derivative with respect to $u$.

For the scalar field we thus expect

\begin{eqnarray}
\dot{M}_B(u) &=\;-\;\frac{1}{2\,\sqrt{\pi}}\;\oint\left[\dot{\sigma}^0\,\dot{\bar{\sigma}}{}^0\;+\;k\,\dot{\phi}^0\,\dot{\bar{\phi}}{}^0\right] \d S,
\end{eqnarray}

\noindent where $k$ is a positive constant factor and $\phi^0$ is the radiative part of the scalar field, i.e. $\phi = \phi^0 \,r^{-1} + {\cal O}(r^{-2})$. We shall now calculate the Bondi mass for the scalar field which will imply the exact formula for the mass-loss.

\subsection{Massless-KG field}

To compute the Bondi mass we use a method based on the asymptotic twistor equation as described in \cite{JS}. More detail can be found in \cite{HT} or \cite{PR}. In this approach we have to find the asymptotic solution of the Einstein-massless-KG equations in the neighbourhood of $\scri^+$ (in the physical spacetime). We give enough of this for our present purposes in Appendix C.

The Bondi mass is then given by the coefficient $\mu^{(2)}$, which is the $\ord{\Omega^2}$ term in the expansion of the spin coefficient $\mu$. This term is given by (\ref{Exps2}) and reads

\[
\mu^{(2)} =\;-\,\eth\eth\sbar \;-\;\Psi_2^{(0)}\;-\;2\,\Lambda^{(0)}\;-\;\s\,\dsbar.
\]

\noindent Since the term $\eth\eth\bar{\sigma}{}^{(0)}$ vanishes on integration, we find the Bondi mass to be (with the normalization used in I)

\begin{eqnarray}
M_B(u) &=\;-\,\frac{1}{2\sqrt{\pi}}\oint\d S\,\left[ \Psi_2^{(0)}\,+\,2\,\Lambda^{(0)} \,+\,\sigma^{(0)}\,\dsbar \right].
\end{eqnarray}

\noindent Using the expansion of $\Lambda$ given by (\ref{Exps3}) leads to the final expression

\begin{eqnarray}
M_B(u) &=\;-\,\frac{1}{2\sqrt{\pi}}\oint\d S\,\left[ \Psi_2^{(0)}\,+\,\frac{1}{3}\,\partial_u\left(\phi^{(0)}\,\bar{\phi}^{(0)}\right) \,+\,\sigma^{(0)}\,\dsbar \right].
\end{eqnarray}

To find the time derivative of the Bondi mass we use the leading term in the Bianchi identity (\ref{BIB3}):

\begin{eqnarray}
\dot{\Psi}{}_2^{(0)}\;+\;2\,\dot{\Lambda}{}^{(0)} &=\;\eth\Psi_3^{(0)}\,+\,\Phi_{22}^{(0)}\,+\,\sigma^{(0)}\,\Psi_4^{(0)}.
\end{eqnarray}

\noindent The term $\eth\Psi_3^{(0)}$ vanishes on integration.  By (\ref{Exps3}) we have $\Psi_4^{(0)}=-\ddot{\bar{\sigma}}{}^{(0)}$. The leading term of $\Phi_{22}$ is found from (\ref{RicciComps1}) to be $\Phi_{22}^{(0)}= 2 \dot{\phi}{}^{(0)}\dot{\bar{\phi}}{}^{(0)}$, and the mass-loss formula thus acquires the form

\begin{eqnarray}
\dot{M}{}_B(u) &=\;-\,\frac{1}{2\sqrt{\pi}}\oint\d S\,\left[\ds\,\dsbar\;+\;2\,\dot{\phi}{}^{(0)}\,\dot{\bar{\phi}}{}^{(0)}\right].
\end{eqnarray}

\noindent This expression is manifestly non-positive. If we demand the spacetime to be periodic, the Bondi mass must be constant, i.e.

\[
\dsbar \;=\;\dot{\phi}^{(0)}\;=\;0.
\]


\subsection{Conformal-scalar field}

The same calculation can be repeated with minor changes in the case of the conformal-scalar field. Now we obtain following expressions for the Bondi mass and its ``loss":

\begin{eqnarray}\fl
\eqalign{
M_B(u) &=\;-\,\frac{1}{2\sqrt{\pi}}\oint\d S\;\left[ \Psi_2^{(0)}\;+\;\sigma^{(0)}\,\dot{\bar{\sigma}}{}^{(0)}\,\right],\\
\dot{M}{}_B(u) &=\;-\,\frac{1}{2\sqrt{\pi}}\oint\d S\;\left[ \dot{\sigma}^{(0)}\dot{\bar{\sigma}}{}^{(0)}\;
+\;2\,\left(\dot{\phi}{}^{(0)}\right)^2\;-\;\phi^{(0)}\,\ddot{\phi}{}^{(0)}\,\right].
}
\end{eqnarray}

Now the formula for the rate of change of the Bondi mass is not manifestly non-positive, so it can apparently increase as well as decrease. This seems to be a consequence of the fact, that the energy-momentum tensor (\ref{ConfCoupledTab}) does not obey the energy condition $T_{ab}l^a n^b \geq 0$ for arbitrary future null vectors $l^a$ and $n^a$.

However, if the Bondi mass is supposed to be periodic, its overall change $\Delta M_B$ during one period $T$ is non-positive. Indeed,

\begin{eqnarray}
\eqalign{
\Delta M_B &=\;-\,\frac{1}{2\sqrt{\pi}}\,\int\limits_u^{u+T}{\rm d}u\;\oint\d S\;\left[ \ds \dsbar\;+\;3\,\dot{\phi}{}^{(0)}\,\dot{\phi}{}^{(0)}\right]\\
&+\;\frac{1}{2\sqrt{\pi}}\,\oint\,\d S\, \left[\phi^{(0)}\,\dot{\phi}{}^{(0)}\right]_u^{u+T},\label{BondiTmp}}
\end{eqnarray}

\noindent where we have integrated the term containing $\ddot{\phi}{}^{(0)}$ by parts. The second term in (\ref{BondiTmp}) vanishes because of periodicity and we arrive at a manifestly non-positive expression for the loss of mass during one period. Such an expression can be periodic only if it is constant, so we again obtain the condition

\[
\dsbar \;=\;\dot{\phi}^{(0)}\;=\;0.
\]

\section{Periodic solutions are necessarily stationary: proof of the theorems}

\subsection{The massless-KG field}
In this section we prove that all periodic asympotically-flat Einstein-massless-KG spacetimes, analytic near $\scri^-$ in the coordinates we shall introduce, are necessarily stationary. First we set up a coordinate system, choose the null tetrad and fix the conformal gauge as in paper I, and the justification for the assertions below is given there. The coordinates are denoted $x^\mu=(v,r,\theta, \phi)$. Here $v$ is the affine parameter along the generators of $\scri^-$ and has the meaning of the advanced time. The coordinate $r$ is an affine parameter along the null geodesics ingoing from $\scri^-$ with the property $\Omega = r + \ord{r^2}$, and $(\theta, \phi)$ are standard spherical coordinates on the unit sphere. The NP operators  $D, \Delta$ and $\delta$ representing derivatives in the directions of the vectors $l, n$ and $m$ (constituting the null tetrad) can be expressed in coordinates $x^\mu$ in the following way:

\begin{eqnarray}
\eqalign{
D &=\;\partial_v\;-\;H\,\partial_r\;+\;C^I\,\partial_I,\\
\Delta &=\;\partial_r,\\
\delta &=\;P^I\,\partial_I.\label{NullTetrad}}
\end{eqnarray}

\noindent The metric functions $H, C^I$ and $P^I$ are governed by the frame equations

\begin{eqnarray}
\Delta\,H &=  -\,(\eps+\bar{\eps}),\label{FEq1}\\
\delta\,H &=  -\,\kappa,\label{FEq2}\\
\Delta\,C^I &=  -\,2\,\pi\,P^I\;-\;2\,\bar{\pi}\,\bar{P}^I,\label{FEq3}\\
\bar{\delta}\,P^I\;-\;\delta\,\bar{P}^I &=  (\alpha-\bar{\beta})\,P^I\;-\;(\bar{\alpha}-\beta)\,\bar{P}^I,\label{FEq4}\\
\Delta\,P^I &=  -\,(\mu - \gamma + \bar{\gamma})\,P^I\;-\;\bar{\lambda}\,\bar{P}^I,\label{FEq5}\\
\delta\,C^I\;-\;D\,P^I &=  -\,(\rho+\eps-\bar{\eps})\,P^I\;-\;\sigma\,\bar{P}^I,\label{FEq6}
\end{eqnarray}
which can be understood as determining the non-zero spin coefficients. We choose

\begin{eqnarray}
P^2\;=\;\frac{1}{\sqrt{2}}, &\;\;\;P^3\;=\;\frac{i}{\sqrt{2}\,\sin\theta}\;\;\;\;{\rm on}\;\;\scri^-.
\end{eqnarray}

\noindent The metric functions $H$ and $C^I$ vanish on $\scri^-$ by construction, so the operator $D$ reduces to $\partial_v$ there and we have:

\begin{eqnarray}
H\;=\;C^I \;=\;0, \;\;DP^I\;=\;0\;\;\;\;{\rm on}\;\;\scri^-.
\end{eqnarray}
As a consequence of the choice of the coordinates and the tetrad we have

\begin{eqnarray}\fl
\eqalign{
\rho-\bar{\rho}\;=\;\mu-\bar{\mu}\;=\;\nu \;=\;\pi-\alpha-\bar{\beta}\;=\;\bar{\tau}-\beta-\bar{\alpha}\;=0,&\;\;{\rm everywhere},\\
\alpha\;=\;-\,\beta\;=\;-\frac{1}{2\sqrt{2}}\,\cot\theta,\;\;\;\kappa\;=\;0,&\;\;{\rm on}\;\;\scri^-.\label{TetradGauge}\\
}
\end{eqnarray}

\noindent Exploiting the tetrad gauge freedom corresponding to the rotation of $(m, \bar{m})$ we achieve

\begin{eqnarray}
\eqalign{
\gamma &=\;0\;\;\;\;{\rm everywhere},\\
\eps   &=\;0\;\;\;\;{\rm on}\;\;\scri^-.}
\end{eqnarray}

\noindent Using the conformal gauge freedom we set

\begin{eqnarray}
\mu &=\;0\;\;\;\;{\rm everywhere}.
\end{eqnarray}

Recall from (\ref{LambdaOrd}) that the physical scalar curvature is $\ord{\Omega^3}$. Eqs. (\ref{s01})--(\ref{s23}) for the conformal factor then reveal that on $\scri^-$

\begin{eqnarray}
\eqalign{
F\;=\;\rho\;=\;\sigma \;=\;\pi\;=\;\bar{\tau} &=\;0,\\
\Delta S_0\;=\;\Delta S_1\;=\;\Delta S_2\;=\;DS_2 &=\;0.}
\end{eqnarray}

\noindent Eqs. (\ref{F1}) and (\ref{F2}) for derivatives of $F$ imply

\begin{eqnarray}
\Phi_{00}\;=\;\Phi_{01} &=\;0\;\;\;\;{\rm on}\;\;\scri^-.
\end{eqnarray}

We saw in the previous section that the periodicity of the solution requires the constancy of the Bondi mass. This is expressed by the relations

\begin{eqnarray}
\eqalign{
\psi_0\;=\;\Delta\Psi_0 &=\;0,\\
\fii_0\;=\;D\phi &=\;0,}\;\;\;\;{\rm on}\;\;\scri^-.
\end{eqnarray}

\noindent These equations also imply $D\fii_1 = D\fii_{\bar{1}} = 0$ on $\scri^-$, as can be seen from (\ref{CW1}) and (\ref{CommDdelta}).

First we prove that, assuming periodicity, all NP quantities are time-independent on $\scri^-$, i.e. independent of $v$. This follows immediately from the choices made above for all spin coefficients except for $\lambda$. The Ricci identity (\ref{RI7}) and Bianchi identity (\ref{BIB1}) show

\begin{eqnarray}
D\lambda \;=\;\Phi_{20}, &\;\;D \Phi_{02}\;=\;0,\;\;\;\;{\rm on}\;\;\scri^-,\label{DlambdaDPhi20}
\end{eqnarray}

\noindent and therefore

\begin{eqnarray}
D^2\lambda &=\;0.\label{Dlambda}
\end{eqnarray}

\noindent By the same argument as in \cite{GS} and \cite{BST}, we conclude

\begin{eqnarray}
D\lambda &=\;0\;\;\;\;{\rm on}\;\;\scri^-,
\end{eqnarray}

\noindent since the equation (\ref{Dlambda}) has a polynomial solution in $v$, but $\lambda$ can be periodic only if it is constant. Equation (\ref{DlambdaDPhi20}) then gives

\begin{eqnarray}
\Phi_{20} &=\;0\;\;\;\;{\rm on} \;\;\scri^-.\label{Phi20}
\end{eqnarray}
The conformal Bianchi identities (\ref{CBI1}), (\ref{CBI3}), (\ref{CBI5}) and (\ref{CBI7}) on $\scri^-$ simplify to

\begin{eqnarray}
\fl
\eqalign{
D\psi_1 &=\;0,\\
D\psi_2 -\bar{\delta}\psi_1 &=\;-\,2\,\alpha\,\psi_1,\\
D\psi_3 -\bar{\delta}\psi_2 &=\;-\,2\,\lambda\,\psi_1 \;+\;\frac{1}{3}\,\left( \phi D\bar{\fii}_{\b1} + \bar{\phi} D \fii_{\b1} \right),\\
D\psi_4 - \bar{\delta}\psi_3 &=\;-\,3\,\lambda\,\psi_2 \;+\;2\,\alpha\,\left( \psi_3 + \bar{\phi} \fii_{\b1} + \phi \bar{\fii}_{\b1}\right) \;-\;4\,\fii_{\b1}\,\bar{\fii}_{\b1}\;+\;\bar{\phi}\,\bar{\delta}\fii_{\b1} \;+\; \phi\,\bar{\delta}\bar{\fii}_{\b1}.
}
\end{eqnarray}

\noindent Applying $D$ to these equations, we immediately see that

\begin{eqnarray}
D^2\psi_n &=& 0\;\;\;\;{\rm on}\;\;\scri^-
\end{eqnarray}

\noindent for all $n$. By periodicity

\begin{eqnarray}
D\psi_n &=& 0\;\;\;\;{\rm on}\;\;\scri^-,
\end{eqnarray}

\noindent so all components of the Weyl spinor are $v-$independent on $\scri^-$. Because $\psi_n = \Omega^{-1} \Psi_n$, we have

\begin{eqnarray}
D\Delta\Psi_n &=\;0\;\;\;\;{\rm on}\;\;\scri^-.
\end{eqnarray}

Finally, we investigate the behaviour of the remaining components of the Ricci tensor, i.e., $\Phi_{11},\Phi_{12}$, $\Phi_{22}$ and $\Lambda$. The Ricci identity (\ref{RI8}) immediately shows

\begin{eqnarray}
\Lambda &=\;0\;\;\;\;{\rm on}\;\;\scri^-.\label{Lambda=0}
\end{eqnarray}

\noindent Since $\mu$ is identically zero not only on $\scri^-$, but also in its neighbourhood, the Ricci identity (\ref{RI11}) on $\scri^-$ reduces to

\begin{eqnarray}
\Phi_{22} &=\;-\,\lambda\,\bar{\lambda}\;\;\;\;{\rm on}\;\;\scri^-,
\end{eqnarray}

\noindent and therefore

\begin{eqnarray}
D\Phi_{22} &=\;0\;\;\;\;{\rm on}\;\;\scri^-.
\end{eqnarray}

\noindent Applying $D$ on the Ricci identity (\ref{RI18}) leads to

\begin{eqnarray}
D\Phi_{21} &=\;0\;\;\;\;{\rm on}\;\;\scri^-.
\end{eqnarray}

\noindent The spin coefficients $\alpha$ and $\beta$ on $\scri^-$ are given by (\ref{TetradGauge}). Inserting these into the Ricci identity (\ref{RI17}) we find

\begin{eqnarray}
\Phi_{11} &=\;\frac{1}{2}\;\;\;\;{\rm on}\;\;\scri^-,\label{Phi11}
\end{eqnarray}

\noindent so $\Phi_{11}$ is obviously $v-$independent on $\scri^-$.

We have already shown that $\fii_1$ and $\fii_{\b1}$ are $v-$independent on $\scri^-$. Equation (\ref{CW2}) implies

\begin{eqnarray}
D\fii_2\;-\;\delta\fii_{\b1} &=\;(\beta-\bar{\alpha})\fii_{\b1}\;\;\;\;{\rm on}\;\;\scri^-.
\end{eqnarray}

\noindent Applying $D$ and assuming periodicity of the scalar field we conclude

\begin{eqnarray}
D\fii_2 &=\;0\;\;\;\;{\rm on}\;\;\scri^-.
\end{eqnarray}

\noindent Projections (\ref{CW3}) and (\ref{CW4}) of the wave equation and commutator (\ref{CommDeltadelta}) applied to $\phi$ reveals, after differentiating with $D$, that $D\Delta Q = 0$ on $\scri^-$, with

\[
Q\in\{\fii_0, \fii_{\b1},
\fii_1\}.
\]

To show the same for $\Delta\fii_2$ we apply $D\Delta$ to (\ref{CW2}) and obtain

\begin{eqnarray}
D^2\Delta\fii_2+2\phi D\Delta\Lambda\;=\;\fii_1 D\Delta\pi -\fii_2 D\Delta(\eps+\bar{\eps}-\rho).\label{D2DeltaFii2}
\end{eqnarray}

\noindent From Ricci identities (\ref{RI6}), (\ref{RI9}), (\ref{RI12})--(\ref{RI15}) we find that $D\Delta Q = 0$ on $\scri^-$ for

\[
Q\in\{\rho, \pi, \alpha, \sigma, \eps, \beta\}.
\]

\noindent Applying $D\Delta$ to (\ref{RI8}) shows

\[
D\Delta\Lambda \;=\;0\;\;\;\;{\rm on}\;\;\scri^-,
\]

\noindent and thus (\ref{D2DeltaFii2}) implies $D^2\Delta\fii_2=0$ on $\scri^-$, so by periodicity

\begin{eqnarray}
D\Delta\fii_2 &=& 0\;\;\;\;{\rm on}\;\;\scri^-.\label{DDeltafii2}
\end{eqnarray}

\noindent Thus, we have proved the lemma

\begin{lemma}\label{Lemma}
The following quantities vanish on $\scri^-$:
\begin{eqnarray}
\eqalign{
H, C^I, \rho, \sigma, \pi, \tau, \kappa, \eps, S_0, S_1, F, \psi_0, \Phi_{00}, \Phi_{01}, \Phi_{02}, \Lambda, \fii_0, \\
DP^I, D\alpha, D\beta, D\lambda, DS_2, \Delta S_0, \Delta S_1, \Delta S_2,\\
D\fii_1, D\fii_{\b1},D\fii_2,D\Delta\fii_0, D\Delta\fii_1,D\Delta\fii_{\b1},D\Delta\fii_2,\\
D\psi_1, D\psi_2, D\psi_3, D\psi_4, D\Phi_{11}, D\Phi_{12}, D\Phi_{22}.
}
\end{eqnarray}
\end{lemma}

\medskip

 Now we set up an induction, with inductive hypothesis:

\medskip

\noindent\emph{Suppose that $D\Delta^j Q = 0$ on $\scri^-$ for $0\leq j \leq k$ with
\[ Q \in \left\{H, C^I, P^I, \eps, \rho, \sigma, \lambda, \pi, \tau, \kappa, \alpha, \beta, F, \psi_n, \Phi_{mn}, \Lambda\right\}, \]
and for $0\leq j \leq k+1$ with $Q\in \left\{S_m, \fii_m\right\}$.}

\medskip

\noindent This holds for $k=0$ by Lemma \ref{Lemma}, so we need to deduce it for $j=k+1$ from its validity for $j\leq k$. Here we closely follow the procedure we used in \cite{BST}.
Applying $D\Delta^k$ on (\ref{FEq1}) we find

\[
D\Delta^{k+1}H\;=\;-D\Delta^k(\eps+\bar{\eps}),
 \]

\noindent where the r.h.s. vanishes on $\scri^-$ by the inductive hypothesis. By a similar argument we can deduce $D\Delta^{k+1}Q = 0$ on $\scri^-$

\begin{itemize}
\item for $H, C^I$ and $P^I$ from
(\ref{FEq1}), (\ref{FEq3}) and (\ref{FEq5});
\item for $\kappa, \eps, \pi, \tau, \lambda, \beta, \sigma, \rho$ and $\alpha$ from
(\ref{RI3}), (\ref{RI6}), (\ref{RI9})--(\ref{RI15});
\item for $F$ from (\ref{F3}), taking $\tilde{\Phi}_{mn}$ from
(\ref{RicciComps1}) and $\tilde{\Lambda}$ from (\ref{Lambda1}) and (\ref{LambdaOrd});
\item for $\Phi_{00}, \Phi_{20}, \Phi_{01}$ and $\Phi_{21}$ from
(\ref{BIA2}), (\ref{BIA4}), (\ref{BIB2}) and (\ref{BIB4});
\item for $\Lambda$, $\Phi_{22}$ and $\Phi_{11}$ from
(\ref{RI8}), (\ref{RI11}) and (\ref{RI17});
\item for $\psi_0, \psi_1, \psi_2$ and $\psi_3$ from
(\ref{CBI2}), (\ref{CBI4}), (\ref{CBI6}) and (\ref{CBI8}).
\end{itemize}

Now, all quantities except for $\psi_4$ are proved to satisfy $D\Delta^{k+1}Q=0$ on $\scri^-$. Applying $D\Delta^{k+1}$ on (\ref{s02}), (\ref{s12}), (\ref{s22}), (\ref{CW3}), (\ref{CW4}) and (\ref{CommDeltadelta}) shows $D\Delta^{k+2}Q = 0$ on $\scri^-$ for $Q\in\{S_0, S_1, S_2, \fii_0, \fii_{\b1}, \fii_1\}$.

Ricci identities (\ref{RI6}), (\ref{RI9}), (\ref{RI12})--(\ref{RI15}) and (\ref{RI8}) in this order imply $D\Delta^{k+2}Q=0$ on $\scri^-$ for

\[
Q\in\{\eps, \pi, \beta, \sigma, \rho, \alpha, \Lambda\}.
\]

\noindent Applying $D\Delta^{k+2}$ on (\ref{CW2}) implies $D^2\Delta^{k+2}\fii_2=0$ on $\scri^-$ and using periodicity we obtain $D\Delta^{k+2}\fii_2=0$ on $\scri^-$.

Finally, acting by $D\Delta^{k+1}$ on (\ref{CBI7}) we find $D^2\Delta^{k+1}\psi_4=0$ on $\scri^-$, and therefore, by periodicity, \[D\Delta^{k+1}\psi_4 = 0\;\;\;\;{\rm on}\;\;\scri^-.\]

\noindent This completes the induction.

We have thus proved that all variables are $v-$independent on $\scri^-$ and, assuming analyticity in $r$, in a finite neighbourhood. Since our set of variables includes also the functions $H, C^I$ and $P^I$ constituting the components of metric tensor, we can conclude that $K=\partial_v$ is a Killing vector of the unphysical metric. However, the conformal factor is $v-$independent as well, so the Lie derivative of the physical metric is

\[{\cal L}_K\tilde{g}_{ab}\;=\;-\,2\,\Omega^{-3}\,g_{ab}\,{\cal L}_K\Omega\;=\;0,\]

\noindent i.e. $K$ is also a Killing vector of the physical metric.

The norm of $K$ is given by the component $g_{vv}$ in the coordinates $x^\mu$. The full form of the metric tensor $g_{\mu\nu}$ can be found in I, eq. (34). The norm of $K$ is then

\[ g(K,K) \;=\;g_{vv}\;=\;2 H \,-\,2 \omega\,\bar{\omega},\]

\noindent where $\omega = - C^I R_I$ and $R_I$ are $\ord{1}$ functions (see (33) in I). Frame equations (\ref{FEq1}) and (\ref{FEq3}) imply $H, C^I=\ord{r^2}$, the Ricci identity (\ref{RI6}) with (\ref{Phi11}) shows $\eps = -1/2$ on $\scri^-$. From these relations we find the norm of the Killing vector to be

\[
g(K,K) \;=\;2\,r^2\;+\;\ord{r^3}.
\]

\noindent We can see that $K$ is null on $\scri^-$ and time-like in its neighbourhood. Our results are summarized in the following theorem.

\begin{theo}\label{one}
A weakly-asymptotically-simple time-periodic solution of the Einstein-massless-KG field equations which is analytic in a neighbourhood of $\scri^-$ in the coordinates introduced above necessarily has a Killing vector which is time-like in the interior and extends to a translation on $\scri^-$.
\end{theo}

\subsection{The conformal-scalar field}

The proof for the conformal-scalar field is essentially the same as in the case of massless-KG field, the only difference lying in the Bianchi identities. These are not so complicated as in the previous case and we present them in their full form in Appendix B, eqs. (\ref{ConfCpldBI1})--(\ref{ConfCpldBI8}). Our results are summarized in the following theorem.

\begin{theo}\label{two}
A weakly-asymptotically-simple time-periodic solution of the Einstein-conformal-scalar equations which is analytic in a neighbourhood of $\scri^-$ in the coordinates introduced above necessarily has a Killing vector which is time-like in the interior and extends to a translation on $\scri^-$.
\end{theo}


We briefly outline the main steps of the proof. We use the same coordinate system and tetrad, defined by (\ref{NullTetrad}) and (\ref{FEq1})--(\ref{FEq6}), so all consequences of the choice of the gauge remain unchanged. Projections of the wave equation (\ref{CW1})--(\ref{CW4}) and equations for the conformal factor, (\ref{s01})--(\ref{F3}) differ only in the presence of the physical scalar curvature $\tilde{\Lambda}$, which in this case is zero. The form of the Ricci identities does not depend on the type of the matter field. Summa summarum, eqs. (\ref{NullTetrad})--(\ref{Phi20}) hold without change.

Now it is straightforward to see from the conformal Bianchi identities (\ref{ConfCpldBI1})--(\ref{ConfCpldBI4}) that the Weyl scalars $\psi_n$ are $v-$independent on $\scri^-$. Next we return to eqs. (\ref{Lambda=0})--(\ref{DDeltafii2}), which are again valid. So, lemma (\ref{Lemma}) holds.

To finalize the proof we need to repeat the induction. In the previous case, in the inductive hypothesis we assumed that each quantity $Q$ satisfies $D\Delta^j Q = 0$ on $\scri^-$, where $0\leq j\leq k$, and in addition, $D\Delta^{k+1} Q = 0$ on $\scri^-$ for $Q \in \{S_a, \fii_a\}$. This was neccesary, since the Bianchi identities contained derivatives of these fields. In the inductive step we were able to prove $D\Delta^{k+1}Q=0$ on $\scri^-$ for $\psi_n$'s and for all other quantities. Moreover, we proved $D\Delta^{k+2}Q=0$ on $\scri^-$ for $Q \in \{S_a, \fii_a\}$.

In this case, the Bianchi identities actually contain the second derivatives of the fields $S_a$ and $\fii_a$, i.e. the third derivatives of $\Omega$ and $\phi$. This is not a problem, however, as all third derivatives are multiplied by $\Omega$. Therefore terms with problematic $D\Delta^{k+2}$-derivatives vanish on $\scri^-$, and the induction can be repeated without change.

\qed

\section{Inheritance}

In the previous section we proved that if both gravitational and scalar fields are periodic near infinity, the spacetime is stationary there and the scalar field does not depend on time. However, there are examples known in which the gravitational field and its matter source do not share the same symmetries (see Paper I for a longer discussion). The question therefore is, whether a stationary gravitational field can be produced by a time-dependent source. In I we showed that this is not the case with an electromagnetic field - once the spacetime is stationary, the electromagnetic field must be too. Let us briefly recall the idea of the proof.

In the electromagnetic case the components of the physical Ricci spinor have the simple form $\tilde{\Phi}_{mn} = \tilde{\phi}_m \tilde{\bar{\phi}}_n$. It is clear, that if the Ricci spinor is to be stationary, the electromagnetic field can depend on time only through the phase of $\tilde{\phi}_m$, i.e. $\tilde{\phi}_m = \tilde{\fii}_m e^{i\chi}$, where $\chi=\chi(v, r, x^I)$, but the modulus $\tilde{\fii}$ is time-independent. Now, by the Bondi mass-loss formula $\phi_0=0$ on $\scri^-$. Using Maxwell's equations we deduced that $\phi_m$ are time-independent on $\scri^-$, and by induction also in its neighbourhood. Therefore, if an asymptotically flat electro-vacuum spacetime is stationary, the electromagnetic field has to inherit stationarity.

The situation is more complicated in the case of the massless-KG field, for now it is not obvious what kind of time-dependence of the scalar field $\tilde{\phi}$ is compatible with the stationarity of the spacetime. We first find this time-dependence and then the result follows, using the Bondi mass-loss formula and induction.

\begin{theo}\label{three}
In a stationary, analytic, weakly-asymptotically-simple solution of the Einstein-massless-KG equations with stationarity Killing vector $K=\partial/\partial v$, the physical massless-KG field must take the form $\tilde{\phi}=e^{i\omega v} \tilde{\phi}_0$, where $\partial_v\tilde{\phi}_0=0$. If the metric is analytic in a neighbourhood of $\scri^-$ in the coordinates introduced above then $\tilde{\phi}$ is in fact time-independent.

\end{theo}

For the first part, we use the coordinate system introduced above and assume the stationarity of the spacetime. Hence, $K=\partial_v$ is a Killing vector of the metric and the Lie derivative ${\cal L}_K$ reduces to simple partial derivative with respect to $v$, which will be also denoted by a dot. Since $\tilde{\Lambda}$ is stationary, Einstein's equations (\ref{EinstEqs1}) imply

\[
\partial_v(\tilde{\fii}_c\tilde{\bar{\fii}}^c)\;=\;0.
\]

\noindent The Lie derivative of the energy-momentum tensor (\ref{KG-Tab}) is then

\[
4\pi{\cal L}_K \widetilde{T}_{ab} \;=\;\tilde{\psi}_a \tilde{\bar{\fii}}_b + \tilde{\fii}_a \tilde{\bar{\psi}}_b + \tilde{\fii}_b \tilde{\bar{\psi}}_a + \tilde{\psi}_b \tilde{\bar{\fii}}_a,
\]

\noindent where $\tilde{\psi}=\dot{\tilde{\phi}}$ and $\tilde{\psi}_a = \nabla_a \tilde{\psi}$. Let us decompose the fields $\tilde{\phi}$ and $\tilde{\psi}$ into real and imaginary parts,

\begin{eqnarray}
\tilde{\phi}\;=\;X + i Y, \;\;\;\; \tilde{\psi}\;=\;U + i W,\label{PhiPsi}
\end{eqnarray}

\noindent with $X,Y, U$ and $W$ being real functions. In this notation we have

\begin{eqnarray}
2 \pi \dot{\tilde{T}}_{ab} &=\;X_a U_b + Y_a W_b + X_b U_a + Y_b W_a\;=\;0, \label{Tabdot}
\end{eqnarray}
where $X_a=\nabla_a X$ etc.

We first consider the case when the gradient fields $X_a$ and $Y_a$ are proportional in some finite region so that, by analyticity, they are proportional everywhere. Thus $X$ and $Y$ are functionally dependent. If either
were constant then that constant would be zero, since $\tilde{\phi}=0$ at
infinity. Thus we may
suppose that $Y$ is a function of $X$ and then
\[\tilde{\fii}_{a}=(1+iY')X_{a},\;\;\tilde{\Box} \tilde{\phi}=(1+iY')\tilde{\Box}
X+iY''\tilde{g}^{ab}X_{a}X_{b},\]
where the prime indicates derivative w.r.t $X$.

If $Y''=0$ then
$Y=aX+b$ for constant $a,b$, but once again $b$ must vanish by asymptotic flatness so
$\tilde{\phi}=(1+ia)X$ and, after rescaling $\tilde{\phi}$ by a constant, we may assume $Y=0$ whence also $W=0$.
Now (\ref{Tabdot}) becomes $X_{(a}U_{b)}=0$ from which necessarily $U_a=0$ and $\dot{\tilde{\phi}}=\,$constant, but then asymptotic flatness forces $\dot{\tilde{\phi}}=0$.

If $Y''\neq 0$ then
\[\tilde{\Box} X=0=\tilde{g}^{ab}X_{a}X_{b}.\]
Now
\begin{equation}\label{set}4\pi \tilde{T}_{ab}=2(1+Y'^2)X_{a}X_{b},\end{equation}
and one may impose on this expression the vanishing of
${\cal{L}}_K \tilde{T}_{ab}$.
Introduce
\[h:={\mathcal{L}}_K X,\]
then this is
\[{\mathcal{L}}_K(4\pi \tilde{T}_{ab})=4Y'Y''hX_aX_b+2(1+Y'^2)(h_aX_b+X_ah_b)=0.\]
For nonzero $h$, this is only possible if $h_a$ is proportional to
$X_a$, so that $h$ is a function of $X$ and this condition becomes
\[4Y'Y''h+4h'(1+Y'^2)=0,\]
which can be integrated to give $h^2(1+Y'^2)=C$, a constant. Now
from (\ref{set})
\[4\pi \tilde{T}_{ab}K^aK^b=2h^2(1+Y'^2)=2C,\]
but this expression must vanish at infinity for asymptotic flatness,
so $C=0$ so $h=0$, and the scalar field inherits the symmetry of the
metric.

When $X_a$ and $Y_a$ are not proportional (except possibly on a set of measure zero), return to (\ref{Tabdot}) and choose a vector field $Z^a$ with $Z^aX_a=0$ but $Z^a Y_a \neq 0$. Contracting (\ref{Tabdot}) with such $Z^a$ we find that $W_b$ is a linear combination of $X_b$ and $Y_b$, from which we deduce

\[
W\;=\;f(X, Y).
\]

\noindent Similarly, contracting (\ref{Tabdot}) with a different $Z^a$ satisfying $Z^a X_a \neq 0$ and $Z^a Y_a = 0$ we arrive at

\[
U\;=\;g(X,Y).
\]

\noindent Inserting this back into (\ref{Tabdot}) we obtain

\[
g_X\, X_a X_b \;+\;2(f_X + g_Y)\, X_{(a} Y_{b)} \;+\; f_Y\,Y_a Y_b\;=\;0,
\]

\noindent where the subscript on $f$ or $g$ indicates the corresponding partial derivative. Since $X_a$ and $Y_a$ are assumed to be linearly independent, all terms in the last equation must vanish separately. We thus have three differential equations for $f$ and $g$. The general solution is

\[
W=f\;=\;\omega X +\beta, \;\;\;\;U=g\;=\;-\omega Y +\gamma,
\]

\noindent with constant $\omega, \beta$ and $\gamma$. Regarding (\ref{PhiPsi}), for the field $\tilde{\psi}$ we have

\[
\tilde{\psi}\;=\;i\omega(X+iY) + (\gamma+i\beta).
\]

\noindent Since $\tilde{\psi} = \dot{\tilde{\phi}}$, we can solve the last equation to find

\begin{eqnarray}
\tilde{\phi} \;=\; \tilde{\phi}_0\, e^{i \omega v}+const. \label{PhiSol}
\end{eqnarray}

\noindent However, the constant second term must be set to zero, as the field itself must vanish at infinity.

We have shown that the most general non-stationary scalar field compatible with stationarity of the spacetime is of the form

\begin{eqnarray}
\tilde{\phi}(v,r,x^I) \;=\; \tilde{\phi}_0(r, x^I)\,e^{i \omega v}.\label{vDep}
\end{eqnarray}
(It is not difficult to show the same result is obtained with a potential term $V(\tilde\phi,\bar{\tilde\phi})$ added as in subsection \ref{ssPot}, with the extra condition that necessarily also $V$ must have the form $V=F(\tilde\phi\bar{\tilde\phi})$.) We shall next show that nonzero $\omega$ leads to the vanishing of $\tilde\phi$. The stationarity of the spacetime implies the constancy of the Bondi mass, so $\fii_0$ is again zero on $\scri^-$. Now, if $\omega=0$, the field $\phi$ is $v-$independent everywhere and is therefore stationary. On the other hand, if $\omega\neq 0$, then expanding $\phi_0$ in the variable $r$ and using (\ref{PhiSol}), we find

\[
i\,\omega\,\phi_0^{(0)}\;=\;0\;\;\;\;{\rm on}\;\;\scri^-.
\]

\noindent so that $\phi_0^{(0)}=0$. Continuing by induction, suppose that $\phi_0^{(j)}=0$ for $0\leq j \leq k$. Acting with $\Delta^k$ on (\ref{CW2}) leads to (recall that $\rho$ and $\eps$ vanish on $\scri^-$)

\[
i\,\omega\,\Delta^{(k+1)}\phi \;=\; 0 \;\;\;\;{\rm on}\;\;\scri^-.
\]

\noindent and since the constant $\omega$ is assumed to be non-zero, it follows immediately that

\[
\phi^{(k+1)}\;=\;0.
\]

\noindent Hence, by induction and analyticity, the field $\phi$ vanishes in a neighbourhood of $\scri^-$. This completes the proof of Theorem \ref{three} and of inheritance for the massless-KG field.

\qed

Let us now turn to the conformal-scalar field. Again, we demand the stationarity of the metric, and therefore also the stationarity of the energy-momentum tensor, but not the stationarity of the scalar field. Unfortunately, the complicated form of the energy-momentum tensor (\ref{ConfCoupledTab}) does not allow us to find the most general time-dependence of $\tilde{\phi}$ compatible with the stationarity of the metric, and thus we cannot proceed as before. In addition, we cannot deduce any concrete condition on $\scri^-$, for we do not have a negative semi-definite mass-loss formula.
Because of these complications we will only show that scalar field inherits the symmetry in a simpler case. Let us consider a complex conformal-scalar field with energy-momentum tensor

\begin{eqnarray}\fl
\tilde{T}_{ab}^{\Complex} &=\;\frac{1}{4\pi}\left[2\,\tilde{\fii}_{(a}\,\tilde{\bar{\fii}}_{b)}\;-\;\frac{1}{2}\,\tilde{g}_{ab}\,\tilde{\fii}_c\,\tilde{\bar{\fii}}^c
\;-\;\frac{1}{2}\,\tilde{\phi}\,\tilde{\nabla}_a\tilde{\bar{\fii}}_b \;-\;\frac{1}{2}\,\tilde{\bar{\phi}}\,\tilde{\nabla}_a\tilde{\fii}_b \;+\; \tilde{\phi}\,\tilde{\bar{\phi}}\,\tilde{\Phi}_{ab}\right].\label{ComplexConfInvTab}
\end{eqnarray}

\noindent The Bondi mass-loss formula (\ref{BondiTmp}) now takes the form

\begin{eqnarray}\fl
\dot{M}_B &=\;-\,\frac{1}{2\sqrt{\pi}}\,\oint \d S\,\left[\ds\,\dsbar\;+\;2\,\dot{\phi}{}^{(0)}\,\dot{\bar{\phi}}{}^{(0)}\;-\;\frac{1}{2}\,\phi^{(0)}\,\ddot{\bar{\phi}}{}^{(0)}\;-\;\frac{1}{2}\,\ddot{\phi}^{(0)}\,\bar{\phi}{}^{(0)}\right].
\label{BondiTMPConfInvComp}
\end{eqnarray}

Although we cannot exclude the existence of some more general time-dependence of $\tilde{\phi}$, for the field of the form (\ref{vDep}) the energy-momentum tensor (\ref{ComplexConfInvTab}) is stationary. In this case we can integrate by parts in (\ref{BondiTMPConfInvComp}) to find as in (\ref{BondiTmp})

\begin{eqnarray}
\Delta{M}_B &=\;-\,\frac{1}{2\sqrt{\pi}}\,\int\limits_v^{v+2\pi/\omega}{\rm d}v\,\oint \d S\,\left[ \ds\,\dsbar\;+\;3\,\dot{\phi}{}^{(0)}\,\dot{\bar{\phi}}{}^{(0)}\right].
\end{eqnarray}

\noindent Since we assume the stationarity of the spacetime, $\ds = 0$. The constancy of the Bondi mass then implies $\dot{\phi}{}^{(0)}=0$, i.e.

\begin{eqnarray}
D \phi\;\equiv\;\fii_0 \;=\;0\;\;\;\;{\rm on}\;\;\scri^-. \label{Dphi=0onscri}
\end{eqnarray}

Now we can proceed as in the case of massless scalar field. By (\ref{Dphi=0onscri}) we have $\omega = 0$ or $\phi^{(0)}=0$. If $\omega=0$, the field is time-independent everywhere. If $\phi^{(0)}=0$, or equivalently, $\phi=0$ on $\scri^-$, we prove by induction that $\phi=0$ everywhere.

\ack P. T. gratefully acknowledges hospitality and financial
support from the Mittag-Leffler Institute, Djursholm, Sweden and the
Charles University, Prague, and useful discussions with Gary Gibbons
and John Stewart. M.S. is grateful to the Mathematical Institute of University of Oxford, and both M.S. and J.B. also thank to Albert Einstein Institute, Golm, for hospitality and support. J.B. acknowledges the discussions with Gary Gibbons and the partial support from the Grant GACR 202/09/00772 of
the Czech Republic, of Grant No LC06014 and MSM0021620860 of the
Ministry of Education. The work of M. S. was supported by the Grant
GAUK no. 22708 of the Charles University and by the Grant GACR-205/09/H033, Czech Republic.
\newpage

\newpage

\section*{Appendix A: The general Newman-Penrose equations}\label{APP-A}
 \setcounter{equation}{0}
 \setcounter{section}{1}
 \setcounter{subsection}{0}
\renewcommand{\theequation}{A\arabic{equation}}
\renewcommand{\thesubsection}{A\arabic{subsection}}
\subsection{Commutation relations}
The operators $D, \Delta, \delta$ and $\bar{\delta}$ satisfy commutation relations

\begin{eqnarray}
D\delta \;-\; \delta D  &=  (\bar{\pi}-\bar{\alpha}-\beta)D - \kappa
\Delta + (\bar{\rho}-\bar{\varepsilon}+\varepsilon)\delta + \sigma
\bar{\delta},\label{CommDdelta}\\
\Delta D  \;-\; D \Delta  &=
(\gamma+\bar{\gamma})D +(\varepsilon + \bar{\varepsilon})\Delta -
(\bar{\tau}+\pi)\delta -
(\tau+\bar{\pi})\bar{\delta},\label{CommDeltaD}\\
\Delta \delta\;-\; \delta \Delta &=  \bar{\nu}D + (\bar{\alpha}+\beta -
\tau)\Delta + (\gamma-\bar{\gamma}-\mu)\delta - \bar{\lambda}
\bar{\delta},\label{CommDeltadelta}\\
\delta\bar{\delta}\; -\; \bar{\delta} \delta
&= (\mu-\bar{\mu})D + (\rho-\bar{\rho})\Delta +
(\bar{\alpha}-\beta)\bar{\delta} -
(\alpha-\bar{\beta})\delta.\label{Commdeltadelta}
\end{eqnarray}

\subsection{Ricci identities}

\NUMPARTS{A}
\begin{eqnarray}
\fl  D \rho - \bar{\delta} \kappa =\rho ^2+\left(\epsilon +\bar{\epsilon
   }\right) \rho -\kappa  \left(3 \alpha +\bar{\beta }-\pi \right)-\tau
   \bar{\kappa }+\sigma  \bar{\sigma }+\Phi_{00},\label{RI1}\\
\fl  D\sigma-\delta\kappa = (\rho+\bar{\rho}+3\eps-\bar{\eps})\sigma - (\tau-\bar{\pi}+\bar{\alpha}+3\beta)\kappa+\Psi_0,\label{RI2}\\
\fl  D\tau-\Delta\kappa = \rho(\tau+\bar{\pi})+\sigma(\bar{\tau}+\pi)+(\eps-\bar{\eps})\tau -(3\gamma+\bar{\gamma})\kappa+\Psi_1+\Phi_{01},\label{RI3}\\
\fl  D\alpha-\bar{\delta}\eps = (\rho +\bar{\eps}-2\eps)\alpha+\beta\bar{\sigma}-\bar{\beta}\eps - \kappa \lambda - \bar{\kappa}\gamma + (\eps+\rho)\pi + \Phi_{10},\label{RI4}\\
\fl  D\beta-\delta\eps = (\alpha+\pi)\sigma + (\bar{\rho}-\bar{\eps})\beta-(\mu+\gamma)\kappa-(\bar{\alpha}-\bar{\pi})\eps + \Psi_1,\label{RI5}\\
\fl  D\gamma-\Delta\eps = (\tau+\bar{\pi})\alpha + (\bar{\tau}+\pi)\beta - (\eps+\bar{\eps})\gamma - (\gamma + \bar{\gamma})\eps + \tau \pi - \nu \kappa\nonumber\\
 +\; \Psi_2 - \Lambda + \Phi_{11},\label{RI6}\\
\fl  D\lambda-\bar{\delta}\pi = (\rho - 3\eps+\bar{\eps})\lambda + \bar{\sigma}\mu + (\pi+\alpha-\bar{\beta})\pi - \nu\bar{\kappa}+\Phi_{20},\label{RI7}\\
\fl  D\mu-\delta\pi = (\bar{\rho}-\eps-\bar{\eps})\mu+\sigma\lambda+ (\bar{\pi}-\bar{\alpha}+\beta)\pi - \nu \kappa + \Psi_2 + 2 \Lambda,\label{RI8}\\
\fl  D\nu-\Delta\pi = (\pi+\bar{\tau})\mu+(\bar{\pi}+\tau)\lambda+(\gamma-\bar{\gamma})\pi - (3\eps+\bar{\eps})\nu+\Psi_3+\Phi_{21},\label{RI9}\\
\fl  \Delta\lambda-\bar{\delta}\nu = -(\mu+\bar{\mu}+3\gamma-\bar{\gamma})\lambda+(3\alpha+\bar{\beta}+\pi-\bar{\tau})\nu-\Psi_4,\label{RI10}\\
\fl  \Delta\mu-\delta\nu = -(\mu+\gamma+\bar{\gamma})\mu-\lambda\bar{\lambda}+\bar{\nu}\pi+(\bar{\alpha}+3\beta-\tau)\nu-\Phi_{22},\label{RI11}\\
\fl  \Delta\beta-\delta\gamma = (\bar{\alpha}+\beta-\tau)\gamma - \mu \tau + \sigma \nu + \eps \bar{\nu} + (\gamma-\bar{\gamma}-\mu)\beta - \alpha\bar{\lambda}-\Phi_{12},\label{RI12}\\
\fl  \Delta\sigma-\delta\tau = -(\mu-3\gamma+\bar{\gamma})\sigma - \bar{\lambda}\rho - (\tau + \beta - \bar{\alpha})\tau + \kappa \bar{\nu}-\Phi_{02},\label{RI13}\\
\fl  \Delta\rho-\bar{\delta}\tau = (\gamma+\bar{\gamma}-\bar{\mu})\rho - \sigma \lambda + (\bar{\beta}-\alpha-\bar{\tau})\tau + \nu \kappa - \Psi_2 - 2 \Lambda,\label{RI14}\\
\fl  \Delta\alpha-\bar{\delta}\gamma = (\rho+\eps)\nu - (\tau+\beta)\lambda + (\bar{\gamma}-\bar{\mu})\alpha + (\bar{\beta}-\bar{\tau})\gamma - \Psi_3,\label{RI15}\\
\fl  \delta\rho-\bar{\delta}\sigma = (\bar{\alpha}+\beta)\rho - (3\alpha-\bar{\beta})\sigma+(\rho-\bar{\rho})\tau+(\mu-\bar{\mu})\kappa -\Psi_1 + \Phi_{01},\label{RI16}\\
\fl  \delta\alpha-\bar{\delta}\beta = \mu\rho-\lambda\sigma + \alpha\bar{\alpha}+\beta\bar{\beta}-2\alpha\beta + (\rho-\bar{\rho})\gamma + (\mu-\bar{\mu})\eps  -  \Psi_2 + \Lambda + \Phi_{11},\label{RI17}\\
\fl  \delta\lambda-\bar{\delta}\mu = (\rho-\bar{\rho})\nu + (\mu-\bar{\mu})\pi + (\alpha+\bar{\beta})\mu+(\bar{\alpha}-3\beta)\lambda-\Psi_3 + \Phi_{21}.\label{RI18}
\end{eqnarray}
\ENDNUMPARTS{A}

\newpage

\subsection{Bianchi identities}

\NUMPARTS{A}
\begin{eqnarray}\eqalign{{\fl}
D\Psi_1-\bar{\delta}\Psi_0-D\Phi_{01}+\delta\Phi_{00} =(\pi - 4 \alpha) \Psi_0+2(2\rho+\varepsilon)\Psi_1-3\kappa\Psi_2+2\kappa\Phi_{11}\\
\
 -\;(\bar{\pi}-2\bar{\alpha}-2\beta)\Phi_{00}-2\sigma\Phi_{10}-
2(\bar{\rho}+\varepsilon)\Phi_{01}+\bar{\kappa}\Phi_{02},\label{BIA1}}\\
\eqalign{
\fl D\Psi_2-\bar{\delta}\Psi_1+\Delta\Phi_{00}-\bar{\delta}\Phi_{01}+2D\Lambda=-\lambda\Psi_0 + 2 (\pi-\alpha)\Psi_1+3\rho \Psi_2-2\kappa\Psi_3\\
 +2\rho\Phi_{11}+\bar{\sigma}\Phi_{02}+\;(2\gamma+2\bar{\gamma}-\bar{\mu})\Phi_{00}-2(\alpha+\bar{\tau})\Phi_{01}-2\tau\Phi_{10},\label{BIA2}}\\
\eqalign{
\fl D\Psi_3-\bar{\delta}\Psi_2-D\Phi_{21}+\delta\Phi_{20}-2\bar{\delta}\Lambda = -2\lambda \Psi_1+3\pi\Psi_2 + 2 (\rho-\varepsilon)\Psi_3-\kappa\Psi_4\\
+2\mu\Phi_{10}-\;2\pi\Phi_{11}-(2\beta+\bar{\pi}-2\bar{\alpha})\Phi_{20}-2(\bar{\rho}-\varepsilon)\Phi_{21}+\bar{\kappa}\Phi_{22},\label{BIA3}}\\
\eqalign{
\fl D\Psi_4-\bar{\delta}\Psi_3+\Delta\Phi_{20}-\bar{\delta}\Phi_{21} =-3\lambda\Psi_2 +2(\alpha+2\pi)\Psi_3+(\rho-4\varepsilon)\Psi_4+2\nu\Phi_{10}\\
-2\lambda\Phi_{11}-\;(2\gamma-2\bar{\gamma}+\bar{\mu})\Phi_{20}-2(\bar{\tau}-\alpha)\Phi_{21}+\bar{\sigma}\Phi_{22},\label{BIA4}}
\end{eqnarray}
\ENDNUMPARTS{A}

\NUMPARTS{A}
\begin{eqnarray}
\eqalign{
\fl \Delta\Psi_0-\delta\Psi_1+D\Phi_{02}-\delta\Phi_{01}=(4\gamma-\mu)\Psi_0-2(2\tau+\beta)\Psi_1+3\sigma\Psi_2\\
+(\bar{\rho}+2\varepsilon-2\bar{\varepsilon})\Phi_{02}+\;2\sigma\Phi_{11}-2\kappa\Phi_{12}-\bar{\lambda}\Phi_{00}+2(\bar{\pi}-\beta)\Phi_{01},\label{BIB1}}\\
\eqalign{
\fl \Delta\Psi_1-\delta\Psi_2-\Delta\Phi_{01}+\bar{\delta}\Phi_{02}-2\delta\Lambda =\nu\Psi_0+2(\gamma-\mu)\Psi_1-3\tau\Psi_2+2\sigma\Psi_3\\
-\bar{\nu}\Phi_{00}+\;2(\bar{\mu}-\gamma)\Phi_{01}+(2\alpha+\bar{\tau}-2\bar{\beta})\Phi_{02}+2\tau\Phi_{11}-2\rho\Phi_{12},\label{BIB2}}\\
\eqalign{
\fl\Delta\Psi_2-\delta\Psi_3+D\Phi_{22}-\delta\Phi_{21}+2\Delta\Lambda=2\nu\Psi_1-3\mu\Psi_2+2(\beta-\tau)\Psi_3+\sigma\Psi_4\\
-2\mu\Phi_{11}-\bar{\lambda}\Phi_{20}+\;2\pi\Phi_{12}+2(\beta+\bar{\pi})\Phi_{21}+(\bar{\rho}-2\varepsilon-2\bar{\varepsilon})\Phi_{22},\label{BIB3}}\\
\eqalign{
\fl\Delta\Psi_3-\delta\Psi_4-\Delta\Phi_{21}+\bar{\delta}\Phi_{22}=3\nu\Psi_2-2(\gamma+2\mu)\Psi_3+(4\beta-\tau)\Psi_4-2\nu\Phi_{11}\\
-\bar{\nu}\Phi_{20}+\;2\lambda\Phi_{12}+2(\gamma+\bar{\mu})\Phi_{21}+(\bar{\tau}-2\bar{\beta}-2\alpha)\Phi_{22},\label{BIB4}}
\end{eqnarray}
\ENDNUMPARTS{A}

\NUMPARTS{A}
\begin{eqnarray}
\fl D\Phi_{11}-\delta\Phi_{10}+\Delta\Phi_{00}-\bar{\delta}\Phi_{01}+3D\Lambda = (2\gamma+2\bar{\gamma}-\mu-\bar{\mu})\Phi_{00}+(\pi-2\alpha-2\bar{\tau})\Phi_{01}\nonumber\\
\fl \hspace{0.5cm} +\;(\bar{\pi}-2\bar{\alpha}-2\tau)\Phi_{10}+2(\rho+\bar{\rho})\Phi_{11}+\bar{\sigma}\Phi_{02}
+\sigma\Phi_{20}-\bar{\kappa}\Phi_{12}-\kappa\Phi_{21},\label{BIC1}\\~\nonumber\\
\eqalign{
\fl D\Phi_{12}-\delta\Phi_{11}+\Delta\Phi_{01}-\bar{\delta}\Phi_{02}+3\delta\Lambda = (2\gamma-\mu-2\bar{\mu})\Phi_{01}+\bar{\nu}\Phi_{00}-\bar{\lambda}\Phi_{10}\\
\fl \hspace{0.5cm}+\;2(\bar{\pi}-\tau)\Phi_{11}+(\pi+2\bar{\beta}-2\alpha-\bar{\tau})\Phi_{02}
+(2\rho+\bar{\rho}-2\bar{\varepsilon})\Phi_{12}+\sigma\Phi_{21}-\kappa\Phi_{22},\label{BIC2}}\\~\nonumber\\
\fl D\Phi_{22}-\delta\Phi_{21}+\Delta\Phi_{11}-\bar{\delta}\Phi_{12}+3\Delta\Lambda = \nu \Phi_{01}+\bar{\nu}\Phi_{10}-2(\mu+\bar{\mu})\Phi_{11}-\lambda\Phi_{02}-\bar{\lambda}\Phi_{20}\nonumber\\
\fl \hspace{0.5cm}+\;(2\pi-\bar{\tau}+2\bar{\beta})\Phi_{12}+(2\beta-\tau+2\bar{\pi})\Phi_{21}
+\;(\rho+\bar{\rho}-2\varepsilon-2\bar{\varepsilon})\Phi_{22}.\label{BIC3}
\end{eqnarray}
\ENDNUMPARTS{A}


\newpage
\section*{Appendix B: The conformal field equations}\label{APP-B}

 \setcounter{section}{1}
 \setcounter{subsection}{0}
\renewcommand{\theequation}{B\arabic{equation}}
\renewcommand{\thesubsection}{B\arabic{subsection}}

\subsection{The conformally-rescaled wave equation}

The wave equation $\widetilde{\Box}\tilde{\phi}=0$ in the physical spacetime is not conformally invariant. If $\tilde{\phi}$ is the solution of the physical wave equation, then the unphysical scalar field $\phi$ satisfies equation (\ref{ConfEqScField}),

\begin{eqnarray}
\nabla_A^{A^\prime}\fii_{BA^\prime} &=\;2\,\left(\Lambda-\Omega^{-2}\tilde{\Lambda}\right)\,\phi\,\epsilon_{AB}.
\end{eqnarray}

\noindent The projections of this equation are the following:

\begin{eqnarray}
\fl
D\fii_1 - \delta\fii_0 &=(\bar{\pi}-\bar{\alpha}-\beta)\fii_0 + (\bar{\rho}+\eps-\bar{\eps})\fii_1 + \sigma \fii_{\b1} - \kappa \fii_2,
\label{CW1}\\
\fl
D\fii_2-\delta\fii_{\b1}  &=-\mu \fii_0 + \pi \fii_1 +(\bar{\pi}-\bar{\alpha} +\beta)\fii_{\b1}+(\bar{\rho}-\eps-\bar{\eps})\fii_2-2\phi(\Lambda-\Omega^{-2}\tilde{\Lambda}),
\label{CW2}\\
\fl
\Delta\fii_0 - \bar{\delta}\fii_1 &=\;(\gamma+\bar{\gamma} - \bar{\mu})\fii_0 + (\bar{\beta}-\alpha-\bar{\tau})\fii_1 - \tau \fii_{\b1} + \rho \fii_2 -2\phi(\Lambda-\Omega^{-2}\tilde{\Lambda}),
\label{CW3}\\
\fl
\Delta\fii_{\b1} - \bar{\delta}\fii_2 &=\;\nu \fii_0 - \lambda \fii_1 + (\bar{\gamma} - \gamma-\bar{\mu})\fii_{\b1} + (\alpha + \bar{\beta} - \bar{\tau}) \fii_2.
\label{CW4}
\end{eqnarray}

\noindent Appropriate equations for the conformal-scalar field can be obtained from (\ref{CW1})--(\ref{CW4}) by setting $\tilde{\Lambda}=0$.

\subsection{Equations for the conformal factor}
The projections of equation (\ref{DDs}),

\begin{eqnarray}
\fl
\nabla_{AA^\prime}s_{BB^\prime} \;=\; \Omega\,\tilde{\Phi}_{ABA^\prime B^\prime}\;-\;\Omega\,\Phi_{ABA^\prime B^\prime}\;+\;\epsilon_{A B}\,\epsilon_{A^\prime B^\prime}\,\left(\Omega\Lambda-\Omega^{-1}\tilde{\Lambda}+F\right),\nonumber
\end{eqnarray}

\noindent are the following\footnote{There is a misprint in Paper I: in eq. (B.2a) the sign at $(\eps+\bar{\eps}) S_0$ should be ``minus" as in  (\ref{s01}). This does not affect the results of Paper I.}:

\begin{eqnarray}
\fl
DS_0 \;-\;(\eps+\bar{\eps})\,S_0\;+\;\bar{\kappa}\,S_1\;+\;\kappa\,\bar{S}_1 \;=\; \Omega\,\tilde{\Phi}_{00}\;-\;\Omega\,\Phi_{00},\;\label{s01}\\
\fl
\Delta S_0\;-\;(\gamma+\bar{\gamma})\,S_0\;+\;\bar{\tau}\,S_1\;+\;\tau\,\bar{S}_1 \;=\; \Omega\,\tilde{\Phi}_{11}\;-\;\Omega\,\Phi_{11}\;+\;\Omega\,\Lambda\;-\;\Omega^{-1}\,\tilde{\Lambda}\;+\;F,\label{s02}\\
\fl
\delta S_0 \;-\; (\bar{\alpha}+\beta)\,S_0 \;+\;\bar{\rho}\,S_1\;+\;\sigma\,\bar{S}_1 \;=\; \Omega\,\tilde{\Phi}_{01}\;-\;\Omega\,\Phi_{01}\label{s03},\\
\fl
DS_1\;-\;\bar{\pi}\,S_0\;+\;(\bar{\eps}-\eps)\,S_1\;+\;\kappa\,S_2 \;=\; \Omega\,\tilde{\Phi}_{01}\;-\;\Omega\,\Phi_{01},\label{s11}\\
\fl
\Delta S_1 \;-\;\bar{\nu}\,S_0\;+\;(\bar{\gamma}-\gamma)\,S_1\;+\;\tau\,S_2 \;=\; \Omega\,\tilde{\Phi}_{12}\;-\;\Omega\,\Phi_{12}\label{s12}\\
\fl
\delta S_1\;-\;\bar{\lambda}\,S_0\;+\;(\bar{\alpha}-\beta)\,S_1\;+\;\sigma\,S_2 \;=\; \Omega\,\tilde{\Phi}_{02}\;-\;\Omega\,\Phi_{02},\label{s13}\\
\fl
\bar{\delta}S_1\;-\;\bar{\mu}\,S_0\;+\;(\bar{\beta}-\alpha)\,S_1\;+\;\rho\,S_2 \;=\; \Omega\,\tilde{\Phi}_{11}\;-\;\Omega\,\Phi_{11}\;-\;\Omega\,\Lambda\;+\;\Omega^{-1}\,\tilde{\Lambda}\;-\;F,\label{s14}\\
\fl
DS_2\;-\;\pi\,S_1\;-\;\bar{\pi}\,\bar{S}_1\;+\;(\eps+\bar{\eps})\,S_2 \;=\; \Omega\,\tilde{\Phi}_{11}\;-\;\Omega\,\Phi_{11}\;+\;\Omega\,\Lambda\;-\;\Omega^{-1}\,\tilde{\Lambda}\;+\;F,\label{s21}\\
\fl
\Delta S_2\;-\;\nu\,S_1\;-\;\bar{\nu}\,\bar{S}_1\;+\;(\gamma+\bar{\gamma})\,S_2 \;=\; \Omega\,\tilde{\Phi}_{22}\;-\;\Omega\,\Phi_{22},\label{s22}\\
\fl
\delta S_2\;-\;\mu\,S_1\;-\;\bar{\lambda}\,\bar{S}_1\;+\;(\bar{\alpha}+\beta)\,S_2 \;=\; \Omega\,\tilde{\Phi}_{12}\;-\;\Omega\,\Phi_{12}.\label{s23}
\end{eqnarray}

\noindent Equation (\ref{FDers}) for the derivatives of $F=(1/2)\Omega^{-1}s_cs^c$,

\begin{eqnarray}
\fl
\nabla_{AA^\prime} F &=& s^{BB^\prime}\,\tilde{\Phi}_{ABA^\prime B^\prime} \;-\;s^{BB^\prime}\,\Phi_{ABA^\prime B^\prime}\;+\;(\Lambda-\Omega^{-2}\,\tilde{\Lambda})\,s_{AA^\prime},
\end{eqnarray}

\noindent has the following projections:

\begin{eqnarray}
\eqalign{
\fl
DF \;= \; S_2\,\tilde{\Phi}_{00}\;-\;S_1\,\tilde{\Phi}_{10}\;+\;S_0\,\tilde{\Phi}_{11}\;-\;\bar{S}_1\,\tilde{\Phi}_{01}\\
-\;S_2\,\Phi_{00}\;+\;S_1\,\Phi_{10}\;-\;S_0\,\Phi_{11}\;+\;\bar{S}_1\,\Phi_{01}\;+\;(\Lambda-\Omega^{-1}\tilde{\Lambda})S_0,\label{F1}
}
\\
\eqalign{
\fl
\delta F \;= \;S_2\,\tilde{\Phi}_{01}\;-\;S_1\,\tilde{\Phi}_{11}\;+\;S_0\,\tilde{\Phi}_{12}\;-\;\bar{S}_1\,\tilde{\Phi}_{02}\\
-\;S_2\,\Phi_{01}\;+\;S_1\,\Phi_{11}\;-\;S_0\,\Phi_{12}\;+\;\bar{S}_1\,\Phi_{02}\;+\;(\Lambda-\Omega^{-1}\tilde{\Lambda})S_1,\label{F2}
}
\\
\eqalign{
\fl
\Delta F \;=\;S_2\,\tilde{\Phi}_{11}\;-\;S_1\,\tilde{\Phi}_{21}\;+\;S_0\,\tilde{\Phi}_{22}\;-\;\bar{S}_1\,\tilde{\Phi}_{12}\\
-\;S_2\,\Phi_{11}\;+\;S_1\,\Phi_{21}\;-\;S_0\,\Phi_{22}\;+\;\bar{S}_1\,\Phi_{12}\;+\;(\Lambda-\Omega^{-1}\tilde{\Lambda})S_2.\label{F3}}
\end{eqnarray}

\subsection{Conformal Bianchi identities for the scalar field}

Let us write the conformal Bianchi identities (\ref{ConfBI0}) for the scalar field in the form

\begin{eqnarray}
X_{ABC A^\prime} &=\;Y_{ABC A^\prime},
\end{eqnarray}

\noindent where (for notation see (\ref{SYMsymbol}))

\begin{eqnarray}
X_{ABC A^\prime} &=\; \nabla^D_{A^\prime}\psi_{ABCD} \;-\; 2\,\phi\,\bar{\phi}\,s^{B^\prime}_{(C}\Phi_{AB)A^\prime B^\prime},\nonumber\\
Y_{ABC A^\prime} &=\;
4\,\SYMM{s\,\fii\,\bar{\fii}}\;+\;2\,\phi\,\SYMM{\nabla\,s\,\bar{\fii}} \;+\;2\,\bar{\phi}\,\SYMM{\nabla\,s\,\fii}\\
&+\;4\,\Omega\,\left[\frac{1}{2}\,\SYMM{\nabla\,\fii\,\bar{\fii}}\;-\;\phi\,\bar{\phi}^2\,\SYMM{s\,\fii\,s}\;-\;\bar{\phi}\,\phi^2\,\SYMM{s\,\bar{\fii}\,s}\right]\\
&-\;4\,\Omega^2\,\phi\,\bar{\phi}\,\SYMM{s\,\fii\,\bar{\fii}}.
\end{eqnarray}

\noindent Since both sides are totally symmetric in $ABC$, we denote their contractions with spinors $o$ and $\iota$ by the number of $\iota$'s in the first index and number of $\bar{\iota}$'s in the second, e.g. $X_{20} = X_{ABC A^\prime} o^A \iota^B \iota^C \bar{o}^{A^\prime},~X_{01} = X_{ABCA^\prime}o^Ao^Bo^C\bar{\iota}^{A^\prime}$.

The components of $X_{ABC A^\prime}$ read as follows:

\begin{eqnarray}
\eqalign{\fl
X_{00} \;=\;-D\psi_1 + \bar{\delta}\psi_0 + (\pi-4\alpha)\psi_0 + 2(\eps+2\rho)\psi_1 - 3\kappa \psi_2 \\
+\phi\bar{\phi}\left(- 2 S_1\Phi_{00} + 2 S_0 \Phi_{01}\right), \label{CBI1}}
\\
\eqalign{\fl
X_{01} \;=\; \Delta\psi_0 - \delta\psi_1 +(\mu - 4\gamma)\psi_0 + 2(\beta+2\tau)\psi_1 - 3 \sigma \psi_2 \\
+ \phi\bar{\phi}\left(2 S_0 \Phi_{02}- 2 S_1 \Phi_{01}\right), \label{CBI2}}\\
\eqalign{\fl
X_{10}\;=\;-D\psi_2+\bar{\delta}\psi_1 - \lambda\psi_0 + 2(\pi-\alpha)\psi_1 + 3\rho\psi_2 - 2 \kappa \psi_3+\\
+(2/3)\phi\bar{\phi}\left( \bar{S}_1\Phi_{01}-S_2\Phi_{00}+2S_0 \Phi_{11} - 2 S_1 \Phi_{10}\right),\label{CBI3} }\\
\eqalign{\fl
X_{11}\;=\;\Delta\psi_1-\delta\psi_2 - \nu \psi_0 + 2(\mu-\gamma)\psi_1+3\tau\psi_2 - 2 \sigma \psi_3 \\
+(2/3)\phi\bar{\phi}\left( \bar{S}_1 \Phi_{02} - S_2 \Phi_{01} + 2 S_0 \Phi_{12} - 2 S_1 \Phi_{11} \right), \label{CBI4}}\\
\eqalign{\fl
X_{20}\;=\;-D\psi_3 + \bar{\delta}\psi_2 - 2 \lambda \psi_1 + 3 \pi \psi_2 + 2 (\rho-\eps) \psi_3 - \kappa \psi_4 \\
+(2/3)\phi\bar{\phi}\left( S_0 \Phi_{21} - S_1 \Phi_{20} + 2 \bar{S}_1 \Phi_{11} - S_2 \Phi_{10} \right), \label{CBI5} }\\
\eqalign{\fl
X_{21}\;=\;\Delta\psi_2 - \delta\psi_3 - 2 \nu \psi_1 + 3 \mu \psi_2 + 2(\tau-\beta) \psi_3 - \sigma \psi_4 \\
+(2/3)\phi\bar{\phi}\left( S_0 \Phi_{22} - S_1 \Phi_{21} + \bar{S}_1 \Phi_{12} - S_2 \Phi_{11} \right), \label{CBI6} }\\
\eqalign{\fl
X_{30}\;=\;-D\psi_4 + \bar{\delta}\psi_3 - 3 \lambda \psi_2 + 2(\alpha+2\pi)\psi_3 + (\rho-4\eps)\psi_4 \\
+ \phi\bar{\phi}\left(2 \bar{S}_1 \Phi_{21} - 2 S_2 \Phi_{20}\right), \label{CBI7} } \\
\eqalign{\fl
X_{31}\;=\;\Delta\psi_3-\delta\psi_4 - 3 \nu \psi_2 + 2(\gamma+2\mu)\psi_3 + (\tau - 4\beta)\psi_4 \\
 + \phi\bar{\phi}\left(2 \bar{S}_1 \Phi_{22} - 2 S_2 \Phi_{21}\right). \label{CBI8} }
\end{eqnarray}

We do not present the projections of $Y_{ABCA^\prime}$ in full detail since they are too long. However, the structure of all the terms entering this spinor allows one to reconstruct its components from a knowledge of the components of spinors $\SYMM{s\fii\bar{\fii}}$ and $\SYMM{\nabla s \fii}$, if appropriate interchanges of $s, \fii$ and $\bar{\fii}$ are made. For expressions of the type $\SYMM{s\fii\bar{\fii}}$ we get

\begin{eqnarray}
\fl
2\,\SYMM{s\,\fii\,\bar{\fii}}_{00}\;=\;2S_1 \fii_0 \bar{\fii}_0 - S_0 (\fii_0 \bar{\fii}_1 + \fii_1 \bar{\fii}_0),\label{sff1}\\
\fl
2\,\SYMM{s\,\fii\,\bar{\fii}}_{01}\;=\;-2S_0 \fii_1 \bar{\fii}_1 + S_1 ( \fii_0 \bar{\fii}_1 +\fii_1 \bar{\fii}_0), \label{sff2}\\
\eqalign{
\fl
6\SYMM{s\,\fii\,\bar{\fii}}_{10}\;=\;-S_0(\fii_0 \bar{\fii}_2 + \fii_2\bar{\fii}_0 + \fii_1 \bar{\fii}_{\b1} + \fii_{\b1} \bar{\fii}_1)+2 S_1 (\fii_0 \bar{\fii}_{\b1} + \fii_{\b1} \bar{\fii}_0) \\
- \bar{S}_1 (\fii_0 \bar{\fii}_1+\fii_1\bar{\fii}_0) + 2 S_2 \fii_0 \bar{\fii}_0,\label{sff3}}\\
\eqalign{
\fl
6\SYMM{s\,\fii\,\bar{\fii}}_{11}\;=\;-2S_0(\fii_2\bar{\fii}_1 + \fii_1 \bar{\fii}_2) + S_1 (\fii_0 \bar{\fii}_2 + \fii_2\bar{\fii}_0 + \fii_1 \bar{\fii}_{\b1} + \fii_{\b1} \bar{\fii}_1) \\
-2\bar{S}_1 \fii_1 \bar{\fii}_1 + S_2 (\fii_0\bar{\fii}_1 + \fii_1 \bar{\fii}_0),\label{sff4}}\\
\eqalign{
\fl
6\SYMM{s\,\fii\,\bar{\fii}}_{20}\;=\;- S_0 (\fii_2 \bar{\fii}_{\b1} + \fii_{\b1} \bar{\fii}_2) + 2 S_1 \fii_{\b1} \bar{\fii}_{\b1} \\
- \bar{S}_1 (\fii_0 \bar{\fii}_2 + \fii_2\bar{\fii}_0 + \fii_1 \bar{\fii}_{\b1} + \fii_{\b1} \bar{\fii}_1) + 2 S_2 (\fii_0 \bar{\fii}_{\b1} + \fii_{\b1} \bar{\fii}_0),\label{sff5}}\\
\eqalign{
\fl
6\SYMM{s\,\fii\,\bar{\fii}}_{21}\;=\;-2S_0 \fii_2 \bar{\fii}_2 + S_1 ( \fii_2 \bar{\fii}_{\b1} + \fii_{\b1} \bar{\fii}_2) \\
- 2\bar{S}_1 (\fii_2 \bar{\fii}_1 + \fii_1 \bar{\fii}_2) + S_2(\fii_0 \bar{\fii}_2 + \fii_2\bar{\fii}_0 + \fii_1 \bar{\fii}_{\b1} + \fii_{\b1} \bar{\fii}_1),\label{sff6}}\\
\fl
2\SYMM{s\,\fii\,\bar{\fii}}_{30}\;=\;- \bar{S}_1 ( \fii_2 \bar{\fii}_{\b1} + \fii_{\b1} \bar{\fii}_2 ) + 2 S_2 \fii_{\b1} \bar{\fii}_{\b1}, \label{sff7}\\
\fl
2\SYMM{s\,\fii\,\bar{\fii}}_{31}\;=\;-2\bar{S}_1 \fii_2 \bar{\fii}_2+S_2(\fii_{\b1} \bar{\fii}_2 + \fii_2 \bar{\fii}_{\b1}).\label{sff8}
\end{eqnarray}

The expressions of type $\SYMM{\nabla s \fii}$ (and their projections) can be slightly simplified by observing that both $s_a$ and $\fii_a$ are gradients of scalar functions, namely $\Omega$ and $\phi$. Since the commutator $\nabla_{A^\prime (A}\nabla^{A^\prime}_{B)}$ annihilates any scalar quantity, we have

\begin{eqnarray}
\nabla_{A^\prime(A}s^{B^\prime}_{B)}\;=\;\nabla_{A^\prime(A}\fii^{B^\prime}_{B)}\;=\;0,
\end{eqnarray}

\noindent and thus

\begin{eqnarray}
\fl
\SYMM{\nabla s \fii}&=\;\frac{1}{2}\,s_{B^\prime(A}\nabla_C^{B^\prime}\fii_{B)A^\prime}\;+\;\frac{1}{2}\,\fii_{B^\prime(A}\nabla_C^{B^\prime}s_{B)A^\prime}\;=\;-\SYMM{s\nabla \fii}-\SYMM{\fii\nabla s}.
\end{eqnarray}

\noindent The components of $\SYMM{s \nabla \fii}$ are

\begin{eqnarray}
\eqalign{
\fl
2\SYMM{s\nabla\fii}_{00}\;=\;S_0 \left[ \delta\fii_0 - (\beta+\bar{\alpha})\fii_0 + \bar{\rho}\fii_1 + \sigma \fii_{\b1}\right]\\
+ S_1\left[ -D\fii_0 + (\eps + \bar{\eps})\fii_0 - \bar{\kappa}\fii_1 - \kappa \fii_{\b1} \right],
\label{snf1}}\\
\eqalign{
\fl
2\SYMM{s\nabla\fii}_{01}\;=\;S_0 \left[ \delta\fii_1 - \bar{\lambda}\fii_0 + \sigma \fii_2 + (\bar{\alpha}-\beta)\fii_1 \right] \\
+S_1 \left[ - D\fii_1 + \bar{\pi}\fii_0 - \kappa \fii_2 + (\eps - \bar{\eps})\fii_1 \right], \label{snf2}}\\
\fl
6\SYMM{s\nabla\fii}_{10}\;=\;S_0 \left[ \Delta\fii_0 + \delta\fii_{\b1}-(\gamma + \bar{\gamma}+\mu)\fii_0 + \bar{\rho}\fii_2 + \bar{\tau}\fii_1 + (\beta+\tau-\bar{\alpha})\fii_{\b1}\right]\nonumber\\
+S_1\left[ -D\fii_{\b1} - \bar{\delta}\fii_0 + (\pi+\alpha+\bar{\beta})\fii_0 - \bar{\kappa} \fii_2 - \bar{\sigma} \fii_1 + (\bar{\eps}-\rho-\eps)\fii_{\b1}\right]\nonumber\\
+\bar{S}_1\left[\delta\fii_0 - (\bar{\alpha}+\beta)\fii_0 + \bar{\rho} \fii_1 + \sigma \fii_{\b1} \right]\label{snf3}\\
+S_2\left[-D\fii_0 + (\eps + \bar{\eps})\fii_0 - \bar{\kappa}\fii_1 - \kappa \fii_{\b1}\right],  \nonumber\\
\fl
6\SYMM{s\nabla\fii}_{11}\;=\;S_0\left[ \Delta\fii_1 + \delta\fii_2 - \bar{\nu}\fii_0 +(\beta+\tau+\bar{\alpha})\fii_2 + (\bar{\gamma}-\gamma-\mu)\fii_1 - \bar{\lambda}\fii_{\b1}\right] \nonumber \\
+S_1\left[ -D\fii_2 - \bar{\delta}\fii_1 + \bar{\mu} \fii_0 - (\rho+\eps+\bar{\eps}) \fii_2 + (\pi + \alpha - \bar{\beta})\fii_1 + \bar{\pi}\fii_{\b1} \right] \nonumber\\
+\bar{S}_1 \left[\delta\fii_1 - \bar{\lambda}\fii_0 + \sigma \fii_2 + (\bar{\alpha}-\beta)\fii_1 \right]\label{snf4} \\
+S_2\left[ - D\fii_1  + \bar{\pi}\fii_0 - \kappa \fii_2 + (\eps-\bar{\eps}) \fii_1 \right],\nonumber\\
\fl
6\SYMM{s\nabla\fii}_{20}\;=\;S_0\left[ \Delta\fii_{\b1} - \nu \fii_0 + \bar{\tau}\fii_2 + (\gamma - \bar{\gamma})\fii_{\b1}\right] \nonumber\\
+S_1 \left[ - \bar{\delta}\fii_{\b1} + \lambda \fii_0 - \bar{\sigma} \fii_2 + (\bar{\beta}-\alpha)\fii_{\b1} \right] \label{snf5} \\
+ \bar{S}_1 \left[ \Delta \fii_0 + \delta\fii_{\b1} - (\gamma + \bar{\gamma}+\mu) \fii_0 + \bar{\rho} \fii_2 + \bar{\tau} \fii_1 + (\beta + \tau - \bar{\alpha}) \fii_{\b1} \right]\nonumber\\
+S_2 \left[ - D\fii_{\b1} - \bar{\delta}\fii_0 + (\bar{\pi}+\alpha + \bar{\beta}) \fii_0 - \bar{\kappa} \fii_2 - \bar{\sigma}\fii_1 + (\bar{\eps} - \eps - \rho)\fii_{\b1}\right],\nonumber\\
\fl
6\SYMM{s\nabla\fii}_{21}\;=\;S_0 \left[ \Delta\fii_2 + (\gamma+\bar{\gamma})\fii_2 - \nu \fii_1 - \bar{\nu} \fii_{\b1} \right], \nonumber\\
+S_1 \left[ - \bar{\delta} \fii_2 - (\alpha + \bar{\beta}) \fii_2 + \lambda \fii_1 + \bar{\mu} \fii_{\b1} \right] \label{snf6} \\
+ \bar{S}_1 \left[ \Delta\fii_1 + \delta \fii_2 - \bar{\nu}\fii_0 + (\beta + \tau + \bar{\alpha}) \fii_2 + (\bar{\gamma}- \gamma - \mu) \fii_1 - \bar{\lambda}\fii_{\b1} \right] \nonumber\\
+ S_2 \left[ - D\fii_2 - \bar{\delta}\fii_1 + \bar{\mu} \fii_0 - (\rho + \eps + \bar{\eps})\fii_2 + (\pi + \alpha - \bar{\beta}) \fii_1 + \bar{\pi}\fii_{\b1} \right],\nonumber\\
\eqalign{
\fl
2\SYMM{s\nabla\fii}_{30}\;=\;\bar{S}_1 \left[ \Delta\fii_{\b1} - \nu \fii_0 + \bar{\tau} \fii_2 + (\gamma-\bar{\gamma})\fii_{\b1} \right] \\
+S_2 \left[ - \bar{\delta}\fii_{\b1} + \lambda \fii_0 - \bar{\sigma} \fii_2 + (\bar{\beta} - \alpha) \fii_{\b1} \right],\label{snf7}}\\
\eqalign{
\fl
2\SYMM{s\nabla\fii}_{31}\;=\;\bar{S}_1 \left[ \Delta\fii_2 + (\gamma+\bar{\gamma}) \fii_2 - \nu \fii_1 - \bar{\nu}\fii_{\b1} \right] \\
+S_2 \left[ - \bar{\delta}\fii_2 - (\alpha+\bar{\beta}) \fii_2 + \lambda \fii_1 + \bar{\mu} \fii_{\b1} \right]. \label{snf8}}
\end{eqnarray}


\subsection{Conformal Bianchi identities for the conformal-scalar field}

The projections of the Bianchi identities (\ref{ConfCoupledBI})

\begin{eqnarray}
\nabla^D_{A^\prime}\,\psi_{ABCD} &=& 3\,s^{B^\prime}_{(C}\,\phi_{AB) A^\prime B^\prime}\;+\;\Omega \,\nabla^{B^\prime}_{(C}\,\phi_{AB)A^\prime B^\prime}
\end{eqnarray}
\nopagebreak
\noindent are as follows:
\nopagebreak

\begin{eqnarray}
\fl D\psi_1\;-\;\bar{\delta}\psi_0 &= (\pi-4\alpha)\psi_0\;+\;2(\eps+2\rho)\psi_1\;-\;3\kappa\psi_2\;-\;3 S_1 \phi_{00}\;+\;3S_0\phi_{01}\nonumber\\ \fl
&+\;\Omega[ D\phi_{01}\;-\;\delta\phi_{00}\;+\;(2\beta -2\bar{\alpha}-\bar{\pi})\phi_{00}\;-\;2(\eps+\bar{\rho})\phi_{01}\;-\;2\sigma\phi_{11}+\bar{\kappa}\phi_{02}]\nonumber,\\ \fl\label{ConfCpldBI1}\\
\fl
D\psi_2\;-\;\bar{\delta}\psi_1 &= -\;\lambda\psi_0\;+\;2(\pi-\alpha)\psi_1\;+\;3\rho\psi_2\;-\;2\kappa\psi_3\nonumber\\ \fl
&+\;2S_0 \phi_{11}\;+\;\bar{S}_1\phi_{01}\;-\;S_2\phi_{00}\;-\;S_1 \phi_{10}\nonumber\\ \fl
&+\;\frac{1}{3}\Omega[2D\phi_{11}-\Delta\phi_{00}+2\delta\phi_{10}-\bar{\delta}\phi_{01}\nonumber\\ \fl
&+\; (2\gamma + 2\bar{\gamma}+2\mu - \bar{\mu})\phi_{00}\;-\;2(\pi+\alpha+\bar{\tau})\phi_{01}\;-\;2(\bar{\pi}+\tau-2\bar{\alpha}) \phi_{10}\nonumber\\ \fl
&+\;2(\rho-2\bar{\rho})\phi_{11}\;+\;2\bar{\kappa}\phi_{12}\;+\;2\kappa\phi_{21}\;+\;\bar{\sigma}\phi_{02}\;-\;2\sigma\phi_{20}]\nonumber,\\ \fl\label{ConfCpldBI2}\\ \fl
D\psi_3\;-\;\bar{\delta}\psi_2 &= -\;2\lambda\psi_1\;+\;2\pi \psi_2\;+\;2(\rho-\eps)\psi_3\;-\;\kappa\psi_4\nonumber\\ \fl
&+\;S_0 \phi_{21}\;-\;S_1 \phi_{20}\;-\;2 S_2 \phi_{10}\;+\;2 \bar{S}_1 \phi_{11}\nonumber\\ \fl
&+\;\frac{1}{3}\Omega[ 2\Delta \phi_{10}\;-\;D\phi_{21}\;-\;2\bar{\delta}\phi_{11}\;+\;\delta\phi_{20}\nonumber\\ \fl
&+\;2\nu\phi_{00}\;-\;2\lambda\phi_{01}\;+\;2(\mu-\bar{\mu}+2\bar{\gamma})\phi_{10}\;-\;2(\pi + 2\bar{\tau})\phi_{11}\nonumber\\ \fl
&+\;2\bar{\sigma}\phi_{12}\;+\;2(\rho-\bar{\rho}+\eps)\phi_{21}\;+\;(2\bar{\alpha}-2\beta-2\tau-\bar{\pi})\phi_{20}\;+\;\bar{\kappa}\phi_{22}]\nonumber,\\ \fl\label{ConfCpldBI3}\\ \fl
D\psi_4\;-\;\bar{\delta}\psi_3 &= -\;3\lambda\psi_2\;+\;2(\alpha+2\pi)\psi_3\;+\;(\rho-4\eps)\psi_4\;+\;3\bar{S}_1\phi_{21}\;-\;3S_2\phi_{20}\nonumber\\ \fl
&+\;\Omega[ \Delta \phi_{20}\;-\;\bar{\delta}\phi_{21}\nonumber\\ \fl
&+\;2\nu\phi_{10}\;-\;2\lambda\phi_{11}\;+\;2(\alpha-\bar{\tau})\phi_{21}\;+\;(2\bar{\gamma}-2\gamma-\bar{\mu})\phi_{20}\;+\;\bar{\sigma}\phi_{22}],\nonumber\\ \fl\label{ConfCpldBI4}\\ \fl
\Delta\psi_0\;-\;\delta\psi_1 &= (4\gamma-\mu)\psi_0\;-\;2(\beta+2\tau)\psi_1\;+\;3\sigma\psi_2\;+\;3 S_1 \phi_{01}\;-\;3 S_0 \phi_{02}\nonumber\\ \fl
&+\;\Omega[-\;D\phi_{02}\;+\;\delta\phi_{01}\nonumber\\ \fl
&-\;\bar{\lambda}\phi_{00}\;+\;2(\bar{\pi}-\beta)\phi_{01}\;+\;2\sigma\phi_{11}\;-\;2\kappa\phi_{12}\;+\;(2\eps-2\bar{\eps}+\bar{\rho})\phi_{02}],\nonumber\\ \fl\label{ConfCpldBI5}\\ \fl
\Delta\psi_1\;-\;\delta\psi_2 &= \nu\psi_0\;+\;2(\gamma-\mu)\psi_1\;-\;3\tau\psi_2\;+\;2\sigma\psi_3\nonumber\\ \fl
&+\;S_2 \phi_{01}\;+\;2S_1\phi_{11}\;-\;2 S_0 \phi_{12}\;-\;\bar{S}_1\phi_{02}\nonumber\\ \fl
&+\;\frac{1}{3}\,\Omega[\Delta\phi_{01}\;-\;\bar{\delta}\phi_{02}\;+\;2\delta\phi_{11}\;-\;2 D \phi_{12}\nonumber\\ \fl
&-\bar{\nu}\phi_{00}\;+\;2(\bar{\mu}-\mu-\gamma)\phi_{01}\;-\;2\bar{\lambda}\phi_{10}\;+\;2(\tau+2\bar{\pi})\phi_{11}\nonumber\\ \fl
&+\;2(\bar{\rho}-\rho-2\bar{\eps})\phi_{12}\;+\;2\sigma\phi_{21}\;+\;(2\pi+2\alpha-2\bar{\beta}+\bar{\tau})\phi_{02}\;-\;2\kappa\phi_{22}]\nonumber,\\ \fl\label{ConfCpldBI6}
\end{eqnarray}

\begin{eqnarray}
\fl \Delta\psi_2\;-\;\delta\psi_3 &= 2\nu\psi_1\;-\;3\mu\psi_2\;+\;2(\beta-\tau)\psi_3\;+\;\sigma\psi_4\nonumber\\ \fl
&+\;2S_2 \phi_{11}\;+\;S_1 \phi_{21}\;-\;S_0 \phi_{22}\;-\;2\bar{S}_1\phi_{12}\nonumber\\ \fl
&+\;\frac{1}{3}\,\Omega[ 2 \Delta\phi_{11}\;+\;\delta\phi_{21}\;-\;2\bar{\delta}\phi_{12}\;-\;D\phi_{22}\nonumber\\ \fl
&-\;2\nu\phi_{01}\;-\;2\bar{\nu}\phi_{10}\;+\;2(2\bar{\mu}-\mu)\phi_{11}\;+\;2(\pi+\bar{\tau}-2\bar{\beta})\phi_{12}\nonumber\\ \fl
&+\;2(\beta+\tau+\bar{\pi})\phi_{21}\;+\;2\lambda\phi_{02}\;-\;\bar{\lambda}\phi_{20}\;+\;(\bar{\rho}-2\rho-2\eps-3\bar{\eps})\phi_{22}]\nonumber,\\ \fl\label{ConfCpldBI7}\\
\fl
\Delta\psi_3\;-\;\delta\psi_4 &= 3\nu\psi_2\;-\;2(\gamma+2\mu)\psi_3\;+\;(4\beta-\tau)\psi_4\;+\;3 S_2 \phi_{21}\;-\;3\bar{S}_1 \phi_{22}\nonumber\\ \fl
&+\;\Omega\,[\Delta\phi_{21}\;-\;\bar{\delta}\phi_{22}\nonumber\\ \fl
&-\;2\nu\phi_{11}\;+\;2\lambda\phi_{12}\;+\;2(\gamma+\bar{\mu})\phi_{21}\;-\;\bar{\nu}\phi_{20}\;+\;(\bar{\tau}-2\alpha-2\bar{\beta})\phi_{22}].\nonumber\\ \fl\label{ConfCpldBI8}
\end{eqnarray}


\newpage
\section*{Appendix C}
 \setcounter{section}{1}
 \setcounter{subsection}{0}
\renewcommand{\theequation}{C\arabic{equation}}
\renewcommand{\thesubsection}{C\arabic{subsection}}

\subsection*{The asympotic solution of the Einstein-massless-scalar-field equations}

Although we want the results at $\scri^-$, we follow the usual convention and find the asymptotic solution of the field equations in the physical spacetime first in the neighbourhood of $\scri^+$. The results can be easily translated to $\scri^-$. For the solution, we closely follow the procedure presented in \cite{JS} for the vacuum spacetimes. The coordinates, tetrad and conformal transformations of the spin basis are identical to those used therein, and in this Appendix, since we don't consider unphysical quantities, we omit the tildes from physical quantities.

We define

\begin{eqnarray}
\fii_0\;=\;D\phi,\;\;\;\fii_1\;=\;\delta\phi,\;\;\;\fii_{\bar{1}}\;=\;\bar{\delta}\phi,\;\;\;\fii_2\;=\;\Delta\phi.
\end{eqnarray}

\noindent The components of the Ricci spinor and the scalar curvature are given by (\ref{RicciComps1}) and (\ref{Lambda1}). The asymptotic behaviour of these quantities is as follows:

\begin{eqnarray}
\eqalign{
\Phi_{00}, \Phi_{01}, \Phi_{02} &=\;{\cal O}(\Omega^4),\\
\Phi_{11}, \Phi_{12},\Lambda &=\; {\cal O}(\Omega^3),\\
\Phi_{22} &=\; {\cal O}(\Omega^2).
}
\end{eqnarray}

Assuming analyticity we can expand any quantity $X = {\cal O}(\Omega^k)$ in a series of the form

\begin{eqnarray}
X &=\; \sum\limits_{i = 0}^\infty X^{(i)}(u, \theta, \phi)\,\Omega^{i+k}.\label{X-expansion}
\end{eqnarray}

\noindent Using the field equations, i.e. the Ricci and Bianchi identities and the frame equations, we arrive at following asymptotic solution for the spin coefficients (setting $\Phi_{mn}=0$ and $\Lambda = 0$ we recover expansions valid for the vacuum case which can be found, e.g. in \cite{JS}, section 3.10):

\begin{eqnarray}
\eqalign{
\sigma &=\;\sigma^{(0)}\,\Omega^2\;+\;\ord{\Omega^4},\\
\rho &=\;-\,\Omega\;+\;\rho^{(2)}\,\Omega^3\;+\;\ord{\Omega^4},\\
\alpha &=\;a\,\Omega\;+\;\alpha^{(1)}\,\Omega^2\;+\;\ord{\Omega^3},\\
\beta &=\;-\,a\,\Omega\;-\;a\,\s\,\Omega^2\;+\;\ord{\Omega^3},\\
\pi &=\;\eth\sbar\,\Omega^2\;+\;\ord{\Omega^3},\\
\lambda &=\;\dsbar\,\Omega\;+\;\lambda^{(2)}\,\Omega^2\;+\;\ord{\Omega^3},\\
\gamma &=\;\gamma^{(2)}\,\Omega^2\;+\;\ord{\Omega^3},\\
\mu &=\;-\,\frac{1}{2}\,\Omega\;+\;\mu^{(2)}\,\Omega^2\;+\;\ord{\Omega^3},\\
\nu &=\;\ord{\Omega}.
\label{Exps1}
}
\end{eqnarray}

\noindent where

\begin{eqnarray}
\eqalign{
\rho^{(2)}&=\;-\left[\sigma^{(0)}\,\bar{\sigma}^{(0)}\,+\,\Phi_{00}^{(0)}\right],\\
a&=\;-\,(2\sqrt{2})^{-1}\,\cot\theta,\\
\alpha^{(1)}&=\;\eth\sbar\,+\,a\,\sbar,\\
\gamma^{(2)} &=\;a\eth\sbar \;-\; a\bar{\eth}\s \;-\;\frac{1}{2}\,\left(\Psi_2^{(0)} \;+\;\Phi_{11}^{(0)}\;-\;\Lambda^{(0)}\right),\\
\lambda^{(2)} &=\;\frac{1}{2}\,\sbar\;-\;\bar{\eth}\eth\sbar,\\
\mu^{(2)} &=\;-\,\eth\eth\sbar \;-\;\Psi_2^{(0)}\;-\;2\,\Lambda^{(0)}\;-\;\s\,\dsbar,\label{Exps2}\\
}
\end{eqnarray}

\noindent For the relevant Weyl scalars and Ricci tensor components we have

\begin{eqnarray}
\eqalign{
\Psi_2 &=\;\Psi_2^{(0)}\,\Omega^3\;+\;\ord{\Omega^4},\\
\Psi_4 &=\;-\,\ddot{\bar{\sigma}}{}^{(0)}\,\Omega\;+\;\ord{\Omega^2},\\
\Phi_{11} &=\;-\,\frac{1}{2}\,\partial_u\left(\phi^{(0)}\,\bar{\phi}{}^{(0)}\right)\,\Omega^3\;+\;\ord{\Omega^4},\\
\Lambda &=\;\frac{1}{6}\,\partial_u\left(\phi^{(0)}\,\bar{\phi}{}^{(0)}\right)\,\Omega^3\;+\;\ord{\Omega^4}.
\label{Exps3}}
\end{eqnarray}


\section*{Appendix D: Selected solutions to the Einstein-conformal-scalar equations}
 \setcounter{section}{1}
 \setcounter{subsection}{0}
\renewcommand{\theequation}{D\arabic{equation}}
\renewcommand{\thesubsection}{D\arabic{subsection}}

We first briefly survey some explicit stationary solutions to the Einstein-conformal-scalar equations which satisfy the requirements of our theorem. To explore the field equation (\ref{EinstEQS2}) further, we also present two families of time-dependent solutions. Some are singular when $\phi^2=1$, some are not and in some $\phi^2$ never takes the value one.

\subsection{Stationary solutions}

Over fifty years ago Buchdahl \cite{BU} demonstrated how from any given static {\it vacuum} solution  a one-parameter family of pairs of solutions of Einstein's equations with massless scalar field can be constructed. Later Bekenstein \cite{bek1} showed how from any Einstein-scalar field solution the corresponding Einstein-\textit{conformal}-scalar field solution can be found. In particular, considering any static vacuum solution in the form

\begin{eqnarray}
\d s^2 &=&  W^2 \d t^2 \;-\;W^{-2}\,h_{ij}\,\d x^i\,\d x^j,\label{AppBekSol1}
\end{eqnarray}

\noindent two Einstein-conformal-scalar solutions are

\begin{eqnarray}
\eqalign{
\d s^2 &= \frac{1}{4}\,\left( W^\beta \pm W^{-\beta}\right)^2 \left[  W^{2\alpha} \d t^2 -W^{-2\alpha} h_{ij}\,\d x^i\,\d x^j\right], \\
\phi &= \sqrt{\frac{3}{4\pi}}\,\frac{1 \mp W^{2\beta}}{1\pm W^{2\beta}},}\label{AppBekSol2}
\end{eqnarray}

\noindent where $\alpha = (1-3\beta^2)^{1/2}$ and $\beta \in \langle -\frac{1}{\sqrt{3}}, \frac{1}{\sqrt{3}}\rangle$ is a free parameter. Upper and lower signs, respectively, in (\ref{AppBekSol2}) correspond to two types of solutions $A$ and $B$. If the solution (\ref{AppBekSol1}) is asymptotically flat, so is the type $A$ solution. Hence, many solutions satisfying our assumptions are available.

A special spherically symmetric solution --after choosing a suitable radial coordinate-- reads

\begin{eqnarray}
\d s^2 &= \left(1- \frac{m}{\bar{r}}\right)^2\,\d t^2\;-\;\left(1-\frac{m}{\bar{r}}\right)^{-2}\d \bar{r}^2 \;-\;\bar{r}^2\,\left( \d \theta^2 + \sin^2\theta \d \phi^2\right), \nonumber\\
\phi &=\sqrt{\frac{3}{4\pi}}\,\frac{m}{\bar{r}-m}.
\end{eqnarray}

The geometry is identical to that of an extreme Reissner-Nordstr\"om  black hole, so it can be analytically continuated to $\bar{r}<m$. However, $\phi$ and $(\nabla_a \phi)(\nabla^a \phi)$ diverge at the ``horizon" $\bar{r}=m$. Nevertheless, this infinite scalar field does not imply an infinite barrier for test scalar charges and the solutions are often regarded as ``black holes with scalar charge" \cite{bek2}. In any case, both geometry and scalar field are analytic at $\bar{r}\rightarrow \infty$ satisfying our requirements.

Bekenstein's work inspired a number of more recent papers: for example, Einstein-conformal-scalar-field solutions were analyzed in arbitrary dimensions \cite{XD}, self-interacting scalar fields were considered \cite{OAF}, and transversable wormholes from massless conformally coupled and other scalar fields non-minimally coupled to gravity were constructed \cite{BV}.

\subsection{FLRW metric}

In this section we present simple homogenous isotropic solutions of the Einstein-conformal-scalar equations. We shall take the metric in the standard form

\begin{eqnarray}
\d s^2 &=\;\d t^2\;-\;a^2(t)\,\left[ \frac{\d r^2}{1-k\,r^2}\;+\;r^2\,\left(\d \theta^2 + \sin^2\theta \,\d \phi^2\right)\right],\label{FLRW}
\end{eqnarray}

\noindent where $k\in\{-1, 0, 1\}$. The energy-momentum tensor is given by (\ref{TabConfInv}). Since this tensor is traceless, the scalar curvature must vanish:

\[
R \;=\; \frac{6}{a^2}\,\left[k\,+\,\dot{a}^2\,+\,\ddot{a}\,a\right]\;=\;0.
\]

\noindent Solutions to this equation are

\begin{eqnarray}
\eqalign{
a(t) &=\;\sqrt{c_1\;-\;k\,(t+c_2)^2}\;\;\;\;{\rm for}\;\;k\neq 0, \\
a(t) &=\;c_1\,\sqrt{2\,t+c_2}\;\;\;\;{\rm for}\;\;k=0.
}
\end{eqnarray}

The conformally-invariant scalar field in the physical spacetime satisfies d'Alembert's equation $\Box\phi=0$. Because the space-time is assumed to be homogeneous and isotropic, we suppose that the field does not depend on the spatial coordinates. D'Alembert's equation then reduces to

\begin{eqnarray}
\Box \phi(t) &=\; \ddot{\phi}\;+\;\frac{3\,\dot{a}\,\dot{\phi}}{a}\;=\;0.\label{FLRWDalemb}
\end{eqnarray}

\noindent For given $a(t)$, the solution can be found explicitly:

\[
\phi(t)\;=\;C_3\;+\;C_4\,\int \frac{\d t}{a^3(t)}.
\]

In the following short discussion we consider three cases for the three possible values of $k$. We solve the wave equation, find the components of the energy-momentum tensor and see that these components do not exhibit the singularity formally present in (\ref{TabConfInv}).

\noindent {\bf a) $\mathbf{k=0}$.} Imposing the initial condition $a(0)=0$ leads to

\begin{eqnarray}
a(t) &=\;\sqrt{2\,C\,t},
\end{eqnarray}

\noindent where $C$ is an arbitrary positive constant. The general solution of (\ref{FLRW}) is

\[
\phi(t) \;=\;\alpha\;+\;\frac{\beta}{\sqrt{t}}.
\]

\noindent Einstein's equations then imply

\[
\alpha \;=\;\pm\,1,
\]

\noindent and $\beta$ is nonzero but arbitrary. Note that $\phi^2=1$ at $t=\beta^2/4$ if $\alpha\beta/|\beta|=-1$, but $\phi^2$ is never one if $\alpha\beta/|\beta|=+1$. The components of the energy momentum-tensor are

\begin{eqnarray}
T_{ab} &=\;\frac{1}{16\,\pi}\,{\rm diag}\left( \frac{3}{2\,t^2}, \frac{C}{t}, \frac{C\,r^2}{t},\frac{C\,r^2\,\sin^2\theta}{t}\right).
\end{eqnarray}

\noindent Obviously, $T_{ab}$ is regular unless $t=0$. This is the expected initial curvature singularity (for example, the Kretschmann invariant $R_{abcd}\,R^{abcd}= 3/(2 t^4)$ diverges for $t=0$). As noted, above the term $(1-\phi^2)$ may or may not vanish depending on the constants of integration but even when it does there is no singularity in $T_{ab}$ despite the form of (\ref{TabConfInv}).

\noindent {\bf b) $\mathbf{k=-1}$.} Again, we demand $a(0)=0$, so $a(t)$ is of the form

\[
a(t)\;=\;\sqrt{t\,(t + C)}.
\]

\noindent The general solution of the wave equation is

\begin{eqnarray}
\phi(t) &=\;\alpha\;+\;\beta\,\frac{2\,t\,+\,C}{2\,a(t)},
\end{eqnarray}

\noindent and Einstein's equations give

\begin{eqnarray}
\alpha\;=\;\cosh\chi, && \beta\;=\;\sinh\chi,
\end{eqnarray}

\noindent where $\chi$ is an arbitrary constant. Now $(1-\phi^2)$ will vanish at some $t>0$ for $\chi<0$ but not for $\chi>0$. The components of the energy-momentum tensor are

\begin{eqnarray}
T_{ab} &=\;\frac{C^2}{32\,\pi\,a^2(t)}\,{\rm diag}\left( \frac{3}{a^2(t)},\frac{1}{1+r^2},r^2, r^2\,\sin^2\theta\right),
\end{eqnarray}

\noindent and they are again singular only for $t=0$, and not at $\phi^2=1$.

\noindent {\bf c) $\mathbf{k=1}$.} Now we impose conditions $a(0)=0$ and $\dot{a}(T)=0$, so that

\[
a(t) \;=\; \sqrt{t\left(2\,T \,-\,t\right)}.
\]

\noindent The solution of the wave equation is

\begin{eqnarray}
\phi(t) &=\; \alpha\;+\;\beta\,\frac{T-t}{a(t)},
\end{eqnarray}

\noindent and Einstein's equations imply

\begin{eqnarray}
\alpha\;=\;\cos\chi, && \beta\;=\;\sin\chi.
\end{eqnarray}

\noindent In this case, $\phi^2$ always takes the value one for some time, but the components of the energy-momentum tensor are

\begin{eqnarray}
T_{ab} &=\;\frac{T^2}{8\,\pi\,a^2(t)}\,{\rm diag}\left( \frac{3}{a^2(t)}, \frac{1}{1-r^2},r^2, r^2\,\sin^2\theta\right),
\end{eqnarray}
and are nonsingular at $\phi^2=1$.

In \cite{bek1}, cosmological solutions were also considered (both conformal scalar field and incoherent radiation), however, singularities in $T_{ab}$ for scalar field were not discussed.


\subsection{PP-waves}

We can find $pp-$wave solutions with this source: consider the $pp-$wave with metric given by

\begin{eqnarray}
\d s^2 &=\;2\,H(u,x,y)\,\d u^2 \;+\; 2\,\d u\,\d v \;-\;\d x^2\;-\;\d y^2.
\end{eqnarray}

\noindent For simplicity, we assume that the scalar field $\phi=\phi(u, x, y)$ does not depend on $v$. The wave equation is then

\begin{eqnarray}
\Box \phi &=\; -\,\phi_{xx}\;-\;\phi_{yy}\;=\;0,
\end{eqnarray}

\noindent with subscripts denoting corresponding derivatives. We can take the general real solution to be

\begin{eqnarray}
\phi(u, x, y) &=\; f(u, x+i y)/2\;+\;f(u, x - i y)/2,
\end{eqnarray}

\noindent where $f$ is an arbitrary real function of two variables. Let us denote

\begin{eqnarray}
K_{ab} &=\; R_{ab} \;+\;8\,\pi\,T_{ab},
\end{eqnarray}

\noindent so that Einstein's equation are $K_{ab}=0$. One of these equations is

\begin{eqnarray}
K_{01} &=\;\frac{\phi_x^2\;+\;\phi_y^2}{1-\phi^2}\;=\;0,
\end{eqnarray}

\noindent from which we find

\begin{eqnarray}
\phi\ &=\;f(u).
\end{eqnarray}

\noindent Then the only remaining non-zero component of $K_{ab}$ is

\begin{eqnarray}
K_{00} &=\;H_{xx} \;+\;H_{yy}\;+\;\frac{2}{1-f^2}\left( f\,f_{uu}\,-\,2 f_u^2\right).
\end{eqnarray}

\noindent Solving the equation $K_{00}=0$ with respect to $H$ we arrive at

\begin{eqnarray}
H &=\;C(u, x + i y)\;+\;C(u, x-i y) \;+\;\frac{x^2 + y^2}{2}\,\frac{\left(2\,f_u^2 \,-\,f\,f_{uu}\right)}{1-f^2}.
\end{eqnarray}

\noindent Here, $C$ and $f$ are arbitrary real functions. As we can now see, the metric function $H$ is singular if ever $f \equiv \phi = \pm 1$ and this, if it occurs, will be a curvature singularity.

In \cite{KK}, a large class of solutions of the Einstein-conformal-scalar equations for colliding plane waves was found by employing the Bekenstein transformation \cite{bek1}.



\newpage
\begin{thebibliography}{99}
\bibitem{BV}
Barcel\'o C. and Visser M. 2000 \emph{Scalar field, energy conditions and transversable wormholes} Class. Quantum Grav. {\bf 17} 3843--3864
\bibitem{bek1}
Bekenstein J. 1974 \emph{Exact solutions of Einstein-conformal scalar equations}
Ann.Phys. {\bf{82}} 535--547
%
\bibitem{bek2}
Bekenstein J. 1975 \emph{Black holes with scalar charge} Ann.Phys. {\bf{91}} 75--82

%
\bibitem{BST}
Bi\v{c}\'ak J., Scholtz M. and Tod P. 2010 \emph{On asymptotically flat solutions of Einstein's equations periodic in time I. Vacuum and electrovacuum solutions} Class. Quantum Grav. {\bf 27} 055007 (24 pp.)
%
\bibitem{BSTII}
Bi\v{c}\'ak J., Scholtz M. and Tod P. 2010 \emph{On asymptotically flat solutions of Einstein's equations periodic in
time II. Spacetimes with scalar-field sources} Class. Quantum Grav. {\bf 27} 175011 (29 pp.)
%
\bibitem{BW}
Bizo\'n P. and Wasserman A. 2000 \emph{On existence of mini-boson stars}  Comm. Math. Phys. {\bf{215}} 357--373.
%
\bibitem{BU}
Buchdahl H. A. 1959 \emph{Reciprocal Static Metrics and Scalar Fields in the General Theory of Relativity} Phys. Rev. {\bf 115} 1325--1328
%
\bibitem{CCJ}
Callan C. G., Coleman S. and Jackiw R. 1970 \emph{A new improved energy-momentum tensor}
Ann. Physics {\bf{59}} 42--73
%
\bibitem{OAF}
Fonarev O. A., 1995 \emph{Exact Einstein scalar field solutions for formation of black holes in a cosmological setting} Class. Quantum Grav. {\bf 12} 1739--1752
%
\bibitem{F}
Friedrich H. 1981 \emph{On the regular and the asymptotic characteristic initial value problem for Einstein's vacuum field equations}
Proc. Roy. Soc. London Ser. {\bf{A 375}} 169--184
%
\bibitem{GS}
Gibbons G. W. and Stewart J. M. 1984 \emph{Absence of asymptotically
flat solutions of Einstein's equations which are periodic and empty
near infinity} in \emph{ Classical general relativity (London, 1983)} 77--94,
Cambridge Univ. Press, Cambridge.
%
\bibitem{hub}
H\"ubner P. 1995 \emph{General relativistic scalar-field models and asymptotic flatness} Class.Quant.Grav. {\bf{12}} 791--808
%
\bibitem{HT}
Hugget S.~A.~and Tod~K.~P. 1985 \emph{An introduction to twistor theory} Cambridge University Press, Cambridge.
%
\bibitem{KK}
Klim\v{c}\'ik~C.~and Koln\'ik~P.~1993 \emph{Interacting Einstein-conformal scalar waves} Phys. Rev. D {\bf 48} 616--621
%
\bibitem{NP}
Newman~E.~T.~and~Penrose R.~1968 \emph{New Conservation Laws for Zero Rest-Mass Fields in Asymptotically Flat Space-Time}
Proc. R. Soc. Lond. {\bf{A 305}} 175--204
%
\bibitem{PR}
Penrose~R.~and~Rindler~W. 1986 \emph{Spinors and space-time} vol II Cambridge Monographs on Mathematical Physics.
 Cambridge University Press, Cambridge.
%
\bibitem{JS}
 Stewart~J.~1990 \emph{Advanced general relativity},
 Cambridge Monographs on Mathematical Physics.
 Cambridge University Press, Cambridge,
%
\bibitem{Winicour}
Winicour~J.~1988 \emph{Massive fields at null infinity}, J. Math. Phys. {\bf 29} 2117--2121
%
\bibitem{XD}
Xanthopoulos B.~C.~ and Dialynas T.~E.~ 1992 \emph{Einstein gravity coupled to a massless conformal scalar field in arbitrary space-time dimensions} J. Math. Phys. {\bf 33} 1463--1471
\end{thebibliography}
\end{document}